\documentstyle{elsart}

\input{epsf}

\begin{document}
\begin{frontmatter}

%\def\thefootnote{\fnsymbol{footnote}}
%\parskip 8pt
%\parindent 1cm
%\topmargin -1cm
%\textheight 23cm
%\textwidth 16cm
%\renewcommand{\arraystretch}{1.2}
%\def\baselinestretch{1.2}
%\oddsidemargin=0.0cm

%\begin{document}

%\thispagestyle{empty}

%\begin{titlepage}
%\begin{center}
%\hfill hep-ph/9911447\\
%\hfill SISSA 138/99/EP\\
\address{~~~~~~~~~~~~~~~~~~~~~~~~~~~~~~~~~~~~~~~~~~~~~~~~~~~~~
~~~~~~~~~~~~~~SISSA 138/99/EP} 

%\hfill{November 23, 1999}\\
%\vskip1cm
\title{Features in the primordial power spectrum of double D-term inflation}
%\end{center}
%\normalsize
%\vskip1cm
%\begin{center}
%\baselineskip=13pt
%
\author{Julien Lesgourgues}
%
%\end{center}
%
%\begin{center}
%\baselineskip=13pt
\address{\it SISSA--ISAS, Via Beirut 2-4, 34013 Trieste, Italy}
%\vglue 0.8cm
%\end{center}

\begin{abstract}
Recently, there has been some interest for building supersymmetric
models of double inflation.  These models, realistic from a particle
physics point of view, predict a broken-scale-invariant power spectrum
of primordial cosmological perturbations, that may explain eventual
non-trivial features in the present matter power spectrum. In previous
works, the primordial spectrum was calculated using analytic slow-roll
approximations. However, these models involve a fast second-order
phase transition during inflation, with a stage of spinodal
instability, and an interruption of slow-roll. For our previous model
of double D-term inflation, we simulate numerically the evolution of
quantum fluctuations, taking into account the spinodal modes, and we
show that the semi-classical approximation can be employed even during
the transition, due to the presence of a second inflaton field. The
primordial power spectrum possesses a rich structure, and possibly, a
non-Gaussian spike on observable scales.
\vskip 0.5cm
{\small
%\hfill hep-ph/9911447 
}

\end{abstract}
%\end{titlepage}

\end{frontmatter}

\section{Introduction}

The standard inflationary prediction \cite{HSGP} concerning the
primordial power spectrum of scalar metric perturbations (denoted
$\Phi$ in the longitudinal gauge) is a simple power-law~:
$k^3<\!|\Phi|^2\!> \;\; \propto \;\;k^{n-1}$. However, for more than ten years,
there has been some interest for models now called
broken-scale-invariant (BSI), predicting deviations from a power-law.
These models generally involve, in addition to the usual inflaton
field, other (effective) fields, driving successive stages of
inflation, or just triggering phase transitions
\cite{S85,KLS,KL,ST,KP,SBB,PS92,P92}. From the beginning, BSI models
have been motivated by observations. After decisive experiments (like
the APM redshift survey and COBE), it appeared that the standard cold
dark matter model (sCDM), even with a power-law primordial spectrum,
was not in agreement with observations. Adding the BSI primordial
spectrum of ``double inflation'' could improve the situation
\cite{GMS,PPS}, but later it was shown that this model cannot account
for small-scale cosmic microwave background (CMB) anisotropy
measurements \cite{LP}. Nevertheless, interest for BSI models revived
recently, when some authors found possible evidence for a feature in
the present matter power spectrum, around 120 $h^{-1}$Mpc
\cite{E,ARS,SG,BJ}~: this is still a controversial and open question,
which future redshift surveys should answer precisely; anyway, in
\cite{LPS1}, it was shown that this feature could be associated with
the BSI step-like spectrum of an inflationary model proposed by
Starobinsky \cite{S92} (combined with a cosmological constant), while
in \cite{SG}, similar conclusions were reached with a spike in the
primordial spectrum. Finally, there has been a recent attempt
\cite{WK} to connect BSI models with a possible deviation from
gaussianity in COBE data \cite{FMG}.

The step-like spectrum of \cite{S92}, based on a non-analyticity in
the inflaton potential, intends to be an effective description of some
underlying, more complicated and more realistic model. As noticed in
\cite{S98}, this underlying model should certainly involve more than
one scalar field, and some rapid phenomenon, such as a phase
transition, occurring approximately 60 or 50 $e$-folds before the end
of inflation. Such models were recently proposed \cite{TS,DD,Ja} in
the context of global supersymmetry or supergravity.\footnote{Also,
while this work was being completed, there has been an interesting
proposal for generating features in the primordial power spectrum
through resonant fermion production during inflation
\cite{CKRT}.} The main advantage of supersymmetric inflationary
models is that they do not require very small parameters, due to the
existence of flat directions in the tree-level potential, and of small
one-loop quantum corrections in the effective potential.

In these works \cite{TS,DD,Ja}, the power spectrum was calculated
using analytic formulas, for modes exiting the Hubble radius during
one of the two slow-roll stages. Indeed, when inflation is driven by a
single slow-rolling scalar field, a simple expression relates the
spectrum of scalar metric perturbations, $k^3 <\!|\Phi|^2\!>$, to the
field potential (and its derivatives) at the time of Hubble crossing
for each mode. As we will see later, in some situations, the analytic
results can be extended to multiple slow-roll inflation. However, in
more complicated situations, it is necessary to calculate the
primordial spectrum from first principles, i.e., to follow numerically
the evolution of the mode functions of field and metric
perturbations~: this amounts in integrating a system of second-order
differential equations. At the end of inflation, the mode functions of
$\Phi$ give the primordial spectrum. Here, we want to perform such a
numerical simulation, in order to check the analytic results, and to
obtain the shape of the primordial spectrum on intermediate scales,
for modes exiting the Hubble radius between the slow-roll stages. For
these scales, and also on larger scales, the calculation appears to be
very interesting, and reveals some unexpected features.

The supersymmetric double inflationary models of refs. \cite{TS,DD,Ja}
are, in the vocabulary of inflation, double hybrid
models,\footnote{Double hybrid or double supersymmetric inflation had
been introduced earlier \cite{TETRA}, not in the context of BSI
primordial spectra, but as a solution to the problem of initial
conditions for hybrid inflation.} involving two inflaton fields, and
two ``trigger'' fields, driving second-order phase transitions~: one
between the two inflationary stages, and one at the end of the last
stage. Since we will focus on scales crossing the Hubble length around
the intermediate phase transition, the second trigger field, the one
ending inflation, will not be important.\footnote{However, the second
phase transition will produce cosmic strings \cite{RJ}, which can be
as important for the formation of structure as the primordial power
spectrum itself \cite{RJ,CHM}. In current studies, these mechanisms
are completely separated, and here we consider the calculation of the
primordial spectrum only. We are grateful to Rachel Jeannerot for
pointing out this fact.} So, our study involves three fields~: the
inflaton field $A$, in slow-roll during both stages of inflation; the
inflaton $B$, in slow-roll during the first stage only, and then
performing damped oscillations; finally, the trigger field $C$, stable
in a local minimum during the first stage, and then performing
oscillations around the true minimum. In fact, $C$ is a complex Higgs
field, charged under a $U(1)$ gauge symmetry, and the phase transition
describes the spontaneous symmetry breaking of this $U(1)$.  The
inflatons $A$ and $B$ are also complex fields, but their phases are
not affected by the evolution, and are fixed from the beginning; so,
we can treat them as real fields.

This fast second-order phase transition, triggered by the
non-slow-rolling field $C$, and taking place approximately 50
$e$-folds before the end of inflation, while the field $A$ is still in
slow-roll, is a new situation for the calculation of the primordial
power spectrum.

Indeed, in the context of inflation, second-order phase transitions
have been first considered by Kofmann and Linde \cite{KL} and Kofmann
and Pogosyan \cite{KP}, but for models with two fields instead of
three~: an inflaton and a trigger field. Then, during a short stage,
corresponding to the beginning of the phase transition, the mass of
the trigger field becomes negative, and adiabatic/isocurvature
perturbations are exponentially amplified. This mechanism is often
called spinodal instability. It results in the formation of a large
and narrow spike in the primordial spectrum, together with the
appearance of topological defects, which are diluted later if the
second stage of inflation is sufficiently long \cite{KL}. These models
were recently studied in more details, in the context of
``supernatural inflation'' \cite{RSG}, and ``natural hybrid
inflation'' \cite{GLW}, and connected with primordial black hole
production. However, with two fields only, there is not as much
freedom as in our model (in particular, because the trigger field must
be light enough for driving a second inflationary stage). Then, the
spike in the primordial spectrum produces enormous density
fluctuations ($\delta \rho / \rho \sim 1$), allowed only on very small
scales; so, the second stage of inflation cannot last more than $30$
$e$-folds. In our model, since we have a third field driving the
second stage of inflation, we will observe for natural parameter
values a moderate spike, which exceeds the approximately flat power
spectrum only by a factor of order one. This is typically the kind of
feature that can improve current fits to the observed matter power
spectrum \cite{SG}, and in contrast with previous models, it is worth
calculating precisely the spike shape and amplitude.

Second-order phase transitions during inflation have also been
recently studied in details by Boyanovsky, de Vega, Holman and
collaborators (see for instance \cite{BVH1,BVH2,CH}). However, in
these works, the Higgs field is in slow-roll, and drives inflation~: it
is a very interesting situation, involving strongly non-linear
evolution, and requiring some non-perturbative approach. The stage of
spinodal instability leads to the emergence of a stochastic
homogeneous background from the coarse-graining of super-Hubble
scales, like in stochastic inflation \cite{SI}. We will see that in
our model, such features are avoided, and the usual perturbative
semi-classical approach can be employed (again, because we have an
inflaton $A$ supporting inflation during and after the transition).

Obviously, the reason for which double hybrid inflationary models did
not get a lot of attention before, is that without supersymmetric
motivations, they appear as quite ``non-minimal'', with many free
parameters (some of them being tuned to unnaturally small values). On
the other hand, if one plays with superpotentials and tries to build
double-inflationary models, double hybrid inflation arises as one of
the simplest possibilities. One can imagine many variants of double
supersymmetric inflation~: the potential can be of the type of double
F-term inflation \cite{TS}, double D-term inflation \cite{DD}, or
mixed; special supergravity effects can arise \cite{Ja}; different
gauge groups can be broken \cite{RJ}; different assumptions can be
made about the field charges. In the following, we will consider only
our previous double D-term model, and neglect supergravity
corrections,\footnote{These corrections are important when the fields
take very large values, close to the Planck mass. In particular, this
may be the case at the beginning of inflation \cite{LR,LT,Ja}. In our
model, with the small gauge coupling constants that we will consider,
supergravity corrections are sub-dominant.} but most of our results
would arise in the other variants, for some (natural) parameter
values.

Which kind of novelty do we expect from this study? 
\begin{itemize}
\item First, since slow-roll conditions are abruptly violated during
the transition (so that the Hubble parameter $H$ will depend as much
on the kinetic energy of the oscillating fields $B$, $C$, as on the
potential energy of the inflaton field $A$), we should devote a
special attention to the evolution of super-Hubble wavelengths during
the intermediate stage. Previous studies \cite{TS,DD,Ja} assume an
analytic expression for the amplitude of large wavelength
perturbations (those which exit the Hubble radius during the first
stage), of the type $k^{3} <\!|\Phi|^2\!> \; \propto \; V \sum_i
(V_i/V_i')^2$ (see the next section for definitions and references),
where the sum runs over slow-rolling fields~: in our case, $A$ and
$B$. Strictly speaking, this formula was originally derived and
employed in the context of multiple slow-roll inflation \cite{S85}. We
want to test this expression in our case; we will see in fact in
section 3 that the violent slow-roll interruption induces a
significant deviation from the analytic prediction (with our choice of
parameters, by a factor 4.3 for $k^{3/2} <\!|\Phi|^2\!>^{1/2}$).

\item second, interesting physics could emerge from the description of
the Higgs field dynamics during the transition. At the very beginning
of the phase transition, the Higgs field has to roll away from an
unstable equilibrium point. For the purpose of the simulation, one may
think that it is sufficient to put by hand an arbitrary small value
for the Higgs zero-mode, and then let this zero-mode roll
down. Actually this is completely misleading. The final result would
depend strongly on the initial small value (which controls the
duration of the spinodal stage). In fact the Higgs dynamics has to be
studied carefully, taking into account, for the Higgs modulus $|C|$,
the exponential amplification of large wavelength modes due to
spinodal instability, and the emergence of a homogeneous background
out of small quantum perturbations. We will see that these features
have a characteristic signature on the primordial spectrum~: a
non-Gaussian spike.
\end{itemize}

Let us now start working. The details of our model and its potential
are given in \cite{DD}. However, a numerical simulation of the
transition would be impossible without a continuous expression for the
one-loop corrections around the critical value $B_c$, at which
symmetry breaking starts. This expression has been calculated for toy
models of F-term and D-term inflation in \cite{JL}, and it is
straightforward to generalize it to our case. The complete effective
potential is given in the Appendix, together with some
remarks on its motivations from string theory. We choose a
particular arbitrary set of parameters, also given in the Appendix.

In section 2., we recall for non-experts the results usually
expected in multiple slow-roll inflation.

In section 3., we treat the problem in the simplest possible way,
i.e., neglecting the Higgs quantum perturbations. Even at this level,
we find that the amplitude of the large-scale plateau exceeds
the analytic prediction.

In section 4., we discuss the Higgs dynamics at the very beginning
of the transition, around the unstable equilibrium point.

In section 5., we go to the next level of complexity for the
primordial spectrum calculation, and take into account the Higgs
longitudinal perturbations (still neglecting the vacuum degeneracy,
which is expected here to be irrelevant). We use the usual
framework of the semi-classical approximation, and carefully justify
this approach.

\section{Well-known results in multiple slow-roll inflation}

Let us recall some well-know results about quantum perturbations in
multiple slow-roll inflation. The matter of this section can be found
in other papers, but we included it for clarifying the results and
notations of the next sections. We follow essentially Polarski \&
Starobinsky \cite{PS} (see also \cite{GW}), and we devote a particular
attention to a subtle problem (the definition of the $V_i$'s), which
is of particular importance in our case.

\subsection{Semi-classical equations}

In absence of a phase transition, there are some well-known techniques
for the calculation of the power spectrum. Usually, the slow-rolling
fields $\varphi_{i,~i=1,..,N}$, are decomposed into a classical
zero-mode plus quantum perturbations~:
\begin{equation} \label{decomp}
\varphi_i ({\bf x},t) = \bar{\varphi}_i (t) + 
\int \frac{d^3 {\bf k}}{(2 \pi)^{3/2}}
\delta \varphi_i ({\bf k},{\bf x},t).
\end{equation}
The field perturbation $\delta \varphi_i$, and the metric
perturbation (in the longitudinal gauge) $\Phi$ can be expanded over an
$N$-dimensional basis of annihilation operators $\hat{a}_j ({\bf
k}),_{~j=1,..,N}$, satisfying canonical commutation relations
$[\hat{a}_j ({\bf k}), \hat{a}^{\dagger}_{j'} ({\bf k})] =\delta_{jj'}
\delta({\bf k}-{\bf k}')$~:
\begin{equation} \label{def.mode.fct}
\delta \varphi_i ({\bf k},{\bf x},t)= \!\!\! \sum_{j=1,..,N} \!\!\! 
e^{ i{\bf k}{\bf x}} \delta \varphi_{ij}  (k,t) \hat{a}_j ({\bf k}) +
e^{-i{\bf k}{\bf x}} \delta \varphi_{ij}^* (k,t) \hat{a}_j^{\dagger} ({\bf k}).
\end{equation}
This expansion defines a set of mode functions $\delta \varphi_{ij}$
and $\Phi_j$ (from now on, we omit all arguments $k$ and $t$).
The basis is chosen in such way that at some initial time, the
perturbation $\delta \varphi_i$ lies along $\hat{a}_i$. 
Commutation relations for the fields yield the Wronskian normalization
conditions for the mode functions, valid at any time~:
\begin{eqnarray} \label{wronskian}
&\sum_{j=1,..,N}& (\delta \varphi_{ij} \delta \dot{\varphi}_{ij}^{*} 
- \delta \varphi^{*}_{ij} \delta \dot{\varphi}_{ij}) 
= i a^{-3},
\nonumber \\
&\sum_{j=1,..,N}& (\Phi_{j} \dot{\Phi}_{j}^* 
- \Phi_{j}^* \dot{\Phi}_{j}) 
= \frac{i}{4 a k^2 m_P^4} \sum_{i=1,..,N} \dot{\bar{\varphi}}_i^2
\end{eqnarray}
(we are using the reduced Planck mass~: $m_P = (8 \pi G)^{-1/2} = 2.4
\times 10^{18}$GeV).  The background equations of motion are~:
\begin{eqnarray}
\ddot{\bar{\varphi}}_i + 3 H \dot{\bar{\varphi}}_i + \frac{\partial
V}{\partial \varphi_i} = 0 &,& \nonumber \\ {\rm
with} \quad H^2 = \frac{1}{3 m_P^2} ( V + \frac{1}{2} \!\sum_{i=1,..,N}
\!\!\!\dot{\bar{\varphi}}_i^2 )& & \label{eq.mot.back}
\end{eqnarray}
(of course, even if not written explicitly, the potential and its
derivatives are evaluated at
$(\bar{\varphi}_1,..,\bar{\varphi}_N)$). They imply the useful
relation $\dot{H} = - \frac{1}{2 m_P^2} \sum_{i}
\dot{\bar{\varphi}}_i^2$.  The perturbation equations of motion read~:
\begin{eqnarray}
\delta \ddot{\varphi}_{ij} + 3H \delta \dot{\varphi}_{ij} + \left(
\frac{k^2}{a^2} + \frac{\partial^2 V}{\partial \varphi_i^2} \right)
\delta \varphi_{ij} &=& 4 \dot{\bar{\varphi}}_i \dot{\Phi}_j - 2
\frac{\partial V}{\partial \varphi_i} \Phi_{j} - \!\!\!\!
\sum_{j'=1,..,N} \!\! \frac{\partial^2 V}{\partial \varphi_i \partial
\varphi_{j'}} \delta \varphi_{j'j}, \nonumber \\ \dot{\Phi}_{j} + H \Phi_{j}
&=& \frac{1}{2 m_P^2} \sum_{i=1,..,N} \dot{\bar{\varphi}}_i \delta
\varphi_{ij}.
\label{eq.mot.pert}
\end{eqnarray}
Initial conditions for the field mode functions inside the Hubble
length (when $k \gg aH$) are given by WKB solutions of equations
(\ref{eq.mot.pert}) (with $\Phi_{j}$ neglected), normalized with
(\ref{wronskian})~:
\begin{equation} \label{WKB1}
\delta \varphi_{ij} = \delta_{ij} 
\frac{a^{-1}}{\sqrt{2 k }} e^{- i \int \frac{k}{a} dt}.
\end{equation}
Equivalent conditions for the metric mode functions are given at the
next order in $a H / k$~:
\begin{equation} \label{WKB1phi}
\Phi_{j} = i \frac{\dot{\bar{\varphi}}_j}{m_P^2 (2 k)^{3/2}} 
e^{- i \int \frac{k}{a} dt}.
\end{equation}
The primordial spectrum is defined as the power spectrum of
$\Phi$, evaluated outside the Hubble radius, during
matter domination~:
\begin{equation}
k^3 <\!0|\; |\Phi|^2 \;|0\!> = k^3 \sum_{j=1,..,N} |\Phi_j|^2.
\label{specprim}
\end{equation}
In fact, in order to obtain the shape of the primordial spectrum, it
is not necessary to follow the evolution of $\Phi$ until matter
domination, but until the end of inflation. Indeed, at this time, all
the modes observable today are outside the Hubble radius, and are
dominated by the so-called ``growing adiabatic solution'', which
evolution is independent of $k$. In other words, the primordial
spectrum $k^3 \sum_{j} |\Phi_j|^2$ changes only by an overall
normalization factor between the end of inflation and matter
domination.

\subsection{Analytic solutions}

We summarize here the analytic solutions for multiple slow-roll
inflation with uncoupled inflaton fields. It is useful to note that in
any case (with or without slow-roll), and for each value of $j$, two
solutions of eqs.(\ref{eq.mot.pert}) have an analytic expression
outside the Hubble radius (when $k \ll a\,H$)~:
\begin{eqnarray}
\frac{\delta \varphi_{ij}}{\bar{\varphi}_i} &=& \frac{1}{a}
\left({\cal C}_j \int_0^t a~dt' - \tilde{\cal C}_j \right), \nonumber \\
\Phi_{j} &=& {\cal C}_{j} \left( 1 - \frac{H}{a} \int_0^t a~dt' \right)
+ \tilde{\cal C}_{j} \frac{H}{a}. \label{exana}
\end{eqnarray}
The mode with coefficient ${\cal C}_{j}$ is the growing adiabatic
mode, the one with $\tilde{\cal C}_j$ the decaying adiabatic mode.
Other solutions ($2N-2$ isocurvature modes for each $j$) have no
generic expressions. There is also a freedom to add to $\Phi_{j}$ an
arbitrary decaying mode $\tilde{\cal D}/a$. If one looks only for
slowly-varying solutions, the system (\ref{eq.mot.pert}) simplifies
in~:
\begin{eqnarray}
3H \delta \dot{\varphi}_{ij} + 
\left( \frac{k^2}{a^2} + \frac{\partial^2 V}{\partial \varphi_i^2} \right)
\delta \varphi_{ij}
&=& 
- 2 \frac{\partial V}{\partial \varphi_i}
\Phi_{j}, \nonumber
\\
H \Phi_{j} &=& \frac{1}{2 m_P^2} \sum_{i}
\dot{\bar{\varphi}}_i \delta \varphi_{ij},
\end{eqnarray}
and admits only $N$ slowly-varying independent solutions (for each
$j$).  Now, during slow-roll stages, we have~:
\begin{equation}
3 H \dot{\bar{\varphi}}_i
+ \frac{\partial V}{\partial \varphi_i} = 0, \qquad H^2 = \frac{V}{3 m_P^2}.
\label{simp.back.eq}
\end{equation}
In this case, the $N$ slowly-varying solutions can be found
analytically \cite{PS}~:
\begin{eqnarray}
\frac{\delta \varphi_{ij}}{\dot{\bar{\varphi}}_i} &=& \frac{{\cal C}_{j}}{H}
- 2 H \left( \frac{\sum_{i'} d_{i'j} V_{i'}}{V} - d_{ij} \right), \nonumber \\
\Phi_{j} &=& - {\cal C}_{j} \frac{\dot{H}}{H^2} 
-H \frac{d}{dt} \left( \frac{\sum_{i'} d_{i'j} V_{i'}}{V} \right). 
\label{solana}
\end{eqnarray}
The solution with coefficient ${\cal C}_{j}$ is the growing adiabatic
mode (as can be seen by comparing with eq.(\ref{exana})) and the modes
with coefficient $d_{ij}$ are the isocurvature modes. For fixed $j$,
only $N-1$ out of the $N$ coefficients $d_{ij}$ are independent, and
the definition is complete only once an arbitrary constraint has been
chosen (for instance, $d_{1j}=0$). The system (\ref{solana}) can be
inverted in~:
\begin{eqnarray}
{\cal C}_{j} &=& - \frac{1}{m_P^2} \sum_{i=1,..,N} \frac{V_i}{\partial
V / \partial \varphi_i} \delta \varphi_{ij}, \nonumber \\ d_{ij} &=&
\frac{\delta \varphi_{ij}}{2 H \bar{\varphi}_i} - \frac{{\cal C}_j}{2
H^2} + \frac{\sum_{i'} d_{i'j} V_{i'}}{V}. \label{invertsys}
\end{eqnarray}
A crucial point, to which we will come back later, is that the
quantities $V_i$ are not uniquely defined. The condition under which
the former solution is valid during slow-roll reads~: $\forall
i,~~d V_i / d t = (\partial V / \partial \varphi_i) \;
\dot{\bar{\varphi}}_i$. So, the $V_i$ can be found by integrating
$\partial V / \partial \varphi_i$ over time or directly over
$\bar{\varphi}_i$, but there is still a freedom to add a constant term
to each $V_i$. This freedom is equivalent to an ambiguity in the
splitting between adiabatic and isocurvature modes. Indeed, we see
from eqs.(\ref{solana}) that adding constants to the $V_i$'s is
equivalent to changing the definition of ${\cal C}_{j}$ and
$d_{ij}$. So, to remove this freedom, we must give a more precise
definition of isocurvature modes during inflation. In fact,
isocurvature modes can be clearly identified at the end of
inflation, from the fact that their contribution to $\Phi_j$ vanishes
\cite{PS,SY}. Using this as a definition, it is possible to calculate
the $V_i$'s in a unique manner, provided that the whole evolution of
background fields is known, till the end of inflation (in contrast
with single-field inflation, for which ${\cal C}_j$ depends only on
background quantities evaluated at $t=t_k$, the time of Hubble
radius crossing).

The most widely-studied multiple inflationary models (for instance,
those with a polynomial potential and no coupling terms) share the
following property.  When one field exits slow-roll, its
time-derivative, which is decreasing, becomes small with respect to
the other ones ($\dot{\bar{\varphi}}_i \rightarrow 0$, $V_i
\rightarrow {\rm cst}$). So, the simplified background equations
(\ref{simp.back.eq}) and the solutions (\ref{solana}) hold
continuously at any time. Simply, at each transition between two
stages, the perturbation equation for the field exiting slow-roll
becomes irrelevant (the perturbation vanishes), but the other
solutions remain valid. In this case, we have a simple prescription
for defining the $V_i$'s.\footnote{I am very grateful to D.~Polarski
and A.~Starobinsky for providing this prescription.} In the last stage
of single-field inflation, driven by one inflaton $\varphi_1$, only
the adiabatic mode contributes to $\Phi_j$, by definition. So, using
eq. (\ref{solana}), we see that we {\it must} have $V_1=V$, and $V_{i
\neq 1} = 0$. This fixes the integration constants~: at any time, we
can obtain the $V_i$ by integrating backward on the field trajectory,
expressed as $(\bar{\varphi}_1,..,\bar{\varphi}_N) \equiv T_i
(\bar{\varphi}_i)$, from a point at the end of the last slow-roll
stage $T_i ( \bar{\varphi}_{i~{\rm end}})$, back to an arbitrary
point~:
\begin{equation}
\label{defvi}
V_i=
\delta_{1i}~~ V(T_i (\bar{\varphi}_{i~{\rm end}}) )
~~+~~
\int_{\bar{\varphi}_{i~{\rm end}}}^{\bar{\varphi}_i} \frac{\partial
V}{\partial \bar{\varphi}_i} ( T_i (\bar{\varphi}_i)) ~~d \bar{\varphi}_i.
\end{equation}
So, at each transition, one potential $V_i$ goes to zero, and in the
last stage only the growing adiabatic mode remains. The interesting
quantity at the end, which is ${\cal C}^2 \equiv \sum_j {\cal
C}_{j}^2$ (since at the end of inflation the primordial spectrum
defined in eq.(\ref{specprim}) is equal to $k^3 \; {\cal C}(k)^2 \; (\dot{H}
/ H^2)^2$), can be immediately found by applying eq.(\ref{invertsys})
at $t_k$. Indeed, at this time, $\delta \varphi_{ij} \simeq
\delta_{ij} \frac{H(t_k)}{\sqrt{2 k^3}}$, and
\begin{equation}
{\cal C}^2 =
\sum_{j=1,..,N} {\cal C}_{j}^2 = \frac{H^2(t_k)}{2 k^3 m_P^4}
\sum_j \left(\frac{V_j}{\partial V / \partial \varphi_j} \right)^2_{t=t_k}. 
\label{slowrollC}
\end{equation}
On the other hand, for models with violent (``waterfall'')
transitions, i.e., for which non-slow-rolling fields acquire large
time-derivatives with respect to $V$, the solution of
eqs.(\ref{solana}) applies separately to different slow-rolling
stages. Between these stages, the background and perturbation
evolution can be quite complicated.  This makes it very difficult to
define properly the $V_i$'s, and to distinguish between the adiabatic
mode and the isocurvature modes (excepted, of course, during the last
slow-rolling stage). So, the standard result of eq.(\ref{slowrollC})
may be difficult to employ, due to the impossibility of defining the
$V_i$'s analytically. The model considered in this paper is an
illustration of this problem, which, to our knowledge, had not been
addressed previously.

\section{Spectrum neglecting Higgs perturbations}

We will now try to calculate the primordial power spectrum
numerically. For a pioneering work on this kind of numerical
simulation, see ref. \cite{SBB}, in which several cases of
multiple field inflation are treated.

The simplest way to study the phase transition is to neglect completely
the Higgs perturbations, 
and deal only with the inflaton perturbations 
$(\delta A_j, \delta B_j)_{j=1,2}$. In other words, we consider $C$ as an exactly homogeneous field~: 
$C=\bar{C} e^{i \bar{\theta}}$, and we solve the following equations~:
\begin{eqnarray}
\ddot{\bar{A}} + 3 H \dot{\bar{A}} + \frac{\partial V}{\partial A}
&=& 0, \qquad {\rm idem~for~} \bar{B},  \bar{C},
\nonumber \\
3 m_P^2 H^2 &=& V + \frac{\dot{\bar{A}}^2 + \dot{\bar{B}}^2 + \dot{\bar{C}}^2}{2},
\nonumber
\\
\ddot{\delta A_j} + 3H \dot{\delta A_j} + 
\left( \frac{k^2}{a^2} + \frac{\partial^2 V}{\partial A^2} \right)
\delta A_j
&=& 4 \dot{\bar{A}} \dot{\Phi}_j - 2 \frac{\partial V}{\partial A}
\Phi_j - \frac{\partial^2 V}{\partial A \partial B} \delta B_j, \nonumber \\
\ddot{\delta B_j} + 3H \dot{\delta B_j} + 
\left( \frac{k^2}{a^2} + \frac{\partial^2 V}{\partial B^2} \right)
\delta B_j
&=& 4 \dot{\bar{B}} \dot{\Phi}_j - 2 \frac{\partial V}{\partial B}
\Phi_j - \frac{\partial^2 V}{\partial A \partial B} \delta A_j, \label{eq.-} \\
\dot{\Phi}_j + H \Phi_j &=& \frac{1}{2 m_P^2} (
\dot{\bar{A}} \delta A_j + \dot{\bar{B}} \delta B_j ).
\nonumber
\end{eqnarray}
Note that the small coupling term $\partial^2 V / \partial A \partial
B$ arises only at the one-loop order, and when $|B|<B_c$. At the
beginning of the phase transition, $\bar{C}$ is sitting at the origin,
which becomes an unstable equilibrium point at the critical $e$-fold
number $N=N_c$. So, we have to ``push'' it away, and introduce an
initial value at $N_c$, invoking quantum fluctuations as a physical
justification. In contrast with the results of section 5, the
simplified calculation of this section does not depend much on this
initial value, provided that it is small enough and does not introduce
a discontinuity.  We take however for $\bar{C}$ the expectation value
of the coarse-grained quantum fluctuations, that will be calculated in
the next section~:
\begin{equation} \label{push}
\bar{C} = < | C_{quantum} |^2>^{1/2} \sim 
\left( \frac{\partial m_c^2}{\partial N} H \right)^{1/3}.
\end{equation}
After $N_c$, $(\bar{B},\bar{C})$ rapidly evolve towards the
equilibrium point $(0, \bar{C}_{eq})$. In fact, it is useful to
distinguish five stages~:
\begin{enumerate}
\item {\bf first slow-roll stage}~: $\bar{A}$ and $\bar{B}$ are in slow-roll,
$\bar{C}=0$.
\item {\bf spinodal stage}~: while $\bar{C}$ grows from the initial
value given above to a critical value $\bar{C}_{spinodal}(\bar{B})$,
the effective mass square $m_c^2$ is negative~: this is spinodal
instability. This stage will have a characteristic signature only in
section 5, when $C$ perturbations will be taken into account. In our
model this stage lasts $\sim 0.15$ $e$-folds.
\item {\bf quasi-static transition}~: $\bar{B}$ and $\bar{C}$ roll down
to the minimum. This evolution is fast with respect to the Universe
expansion ($\sim 0.8$ $e$-folds in our model) and ($\bar{B}$,
$\bar{C}$) are far from being in slow-roll. However, we call this
transition quasi-static, because for the parameters chosen here
(in particular $\beta \simeq 10^{-3}$), $\bar{C}$ remains close to the
valley of its local minima,\footnote{so, we do not deal with the case
in which $\bar{C}$ undergoes a chaotic stage with large
oscillations. This would arise with $10^{-2} \leq \beta \leq 1$. In
this case, we don't have a robust justification for the expression
of one-loop corrections \cite{JL}, and everything becomes more
complicated.} and performs small oscillations around the elliptical
trajectory $\bar{C}^2=\bar{C}_{eq}^2 (1- \bar{B}^2 / \bar{B}_c^2 )$ on
which $\partial V / \partial C=0$.
\item {\bf second stage with oscillations}~: only $\bar{A}$ is in
slow-roll, but the fields ($\bar{B}$, $\bar{C}$) are not stabilized in
their minimum and their oscillations affect the background evolution.
In our model we find that $|\dot{\bar{C}}| \ll |\dot{\bar{A}}|$ at any
time, but, if we average over one oscillation, $|\dot{\bar{B}}| \gg
|\dot{\bar{A}}|$ during approximately 15 $e$-folds.  This stage is
inflationary ($|\dot{H}| \ll H^2$), but differs from a usual slow-roll
stage since the evolution of $H$ is driven by the oscillating field
$\bar{B}$, not the inflaton field $\bar{A}$.
\item {\bf second stage with slow-roll}~:
at some point, $|\dot{\bar{B}}| \ll |\dot{\bar{A}}|$ and we go back to ordinary
single-field slow-roll inflation, during approximately 35 $e$-folds.
\end{enumerate}
The slow-roll analytic prediction for the primordial spectrum is given by eq.
(\ref{slowrollC}), and at first (tree-level) order, the definition
of ($V_A$, $V_B$) resulting from eq.(\ref{defvi}) and from the
potential (eq.(\ref{potential})) is~:
\begin{eqnarray}
V_A &\simeq& V(\bar{A}_{\rm end}) \simeq 
\frac{g_A^2 g_B^2 (\xi_A-\xi_B)^2}{2 (g_A^2 + g_B^2)}, \label{va} \\
V_B &\simeq& \int_0^{\bar{B}} \frac{1}{2} \beta^2 \bar{B} \bar{C}^2 d \bar{B}
\qquad {\rm with}~~ \bar{C}^2=
\max \{ \bar{C}_{eq}^2 (1- \frac{\bar{B}^2}{\bar{B}_c^2} ), 0 \}
\nonumber \\
&=& \left\{
\begin{tabular}{cl}
$\frac{(g_A^2 \xi_A + g_B^2 \xi_B)^2}{2 (g_A^2 + g_B^2)}$
&during the first stage, \\
$\frac{1}{4} \beta^2 \bar{B}^2 \bar{C}_{eq}^2 (1- \frac{\bar{B}^2}{ 2 \bar{B}_c^2} )$
&during the transition, \\
0
&during the second stage. \\
\end{tabular}
\right. \label{vb}
\end{eqnarray} 
Note that at any time $V=V_A+V_B$. This is due to the fact that the
definition (\ref{defvi}) implies $V=V_A+V_B+V_C$, and if $C$ remains
in the valley of minima, $V_C = \int \frac{\partial V}{\partial C} dC
\simeq 0$.

Let us now compare the slow-roll predictions with the results of the
numerical simulation.  We will first discuss in details the results
for two modes~: on with $k < a(t_c) H(t_c)$ and one with $k > a(t_c)
H(t_c)$.

\subsection{Large wavelength results}

We integrate the equations for a mode that crosses the Hubble radius
around $N=-53$, while $N_c=-50$ (the origin for $N$ has been chosen so
that the last inflationary stage ends at $N=0$). The results for
$|\Phi_1|$ and $|\Phi_2|$ are shown on the first plot of
fig.{\ref{fig-}.  Since the final power spectrum will depend on the
final value of $|\Phi_1|$, which is much bigger than $|\Phi_2|$ during
the last stage, it is important to understand the evolution of this
quantity, and to compare with slow-roll predictions.

In first approximation, the evolution of $\Phi_{1}$ can be reproduced
analytically. Indeed, we check that the perturbation $\delta B_{1}$,
though amplified exponentially during the transition, remains a
negligible source term for $\Phi_{1}$ and $\delta A_{1}$, except
during a short period (two $e$-folds) during which it is comparable
with other source terms.  So, at first order, we can just neglect
$\delta B_1$, and deduce the evolution of $\Phi_{1}$ and $\delta
A_{1}$ from a system of two differential equations instead of
three.\footnote{ note that this simplification is appropriate for
finding $\Phi_{1}$ and $\delta A_{1}$, but the exact adiabatic
solution (\ref{exana}) does not satisfy the reduced system (two
equations). This shows that the solution for $\Phi_{1}$ and $\delta
A_{1}$ must be a mixture of adiabatic and isocurvature
perturbations.}. Moreover, we look only for the slowly-varying
solution, which obeys to~:
\begin{equation}
3 H \delta \dot{A}_{1} + \frac{\partial^2 V}{\partial A^2} \delta A_{1}
= -2 \frac{\partial V}{\partial A} \Phi_{1}, \qquad 
H \Phi_{1} = \frac{\dot{\bar{A}} \; \delta A_{1}}{2 m_P^2}.
\end{equation}
There is an exact solution (normalized to $|\delta A_{1}| = \frac{H}{\sqrt{2k^3}}$
at $t=t_k$), that reproduces fairly well the numerical solution~:
\begin{eqnarray}
\delta A_{1} &=& \frac{\dot{\bar{A}}}{\dot{\bar{A}}(t_k)} 
\frac{H}{\sqrt{2 k^3}} \exp \int_{t_k}^t \frac{\dot{\bar{A}}^2}{m_P^2 H} dt,
\nonumber \\
\Phi_{1} &=& \frac{\dot{\bar{A}}^2}{\dot{\bar{A}}(t_k)} 
\frac{1}{2 \sqrt{2 k^3}} \exp \int_{t_k}^t \frac{\dot{\bar{A}}^2}{m_P^2 H} dt.
\label{fairly}
\end{eqnarray}
We must keep in mind that between $t_c$ and some time $t_1$, one has
$\dot{\bar{B}}^2+\dot{\bar{C}}^2 \geq \dot{\bar{A}}^2$ (this is the 
15 $e$-fold ``second stage with oscillations'').
On the other hand, during the
``second stage with slow-roll'', we have 
$2 m_P^2 \dot{H} = - \dot{\bar{A}}^2$, and eqs.(\ref{fairly}) read~: 
\begin{eqnarray}
\delta A_{1} &=& {\cal C}_1 \frac{\dot{\bar{A}}}{H}, \qquad
\Phi_{1} = -{\cal C}_1 \frac{\dot{H}}{H^2}, \nonumber \\
{\cal C}_{1} &=& 
\frac{H(t_k)}{m_P^2 \sqrt{2 k^3}}
\frac{V(t_k)}{\frac{\partial V}{\partial A} (t_k)}
\exp \int_{t_k}^{t} - \frac{\dot{\bar{B}}^2+\dot{\bar{C}}^2}{m_P^2 H} dt.
\end{eqnarray} 
Now, since the contribution to the last integral is made essentially
for $t_c < t < t_1$ with 
$\dot{H} \simeq - \frac{\dot{\bar{B}}^2+\dot{\bar{C}}^2}{2 m_P^2}$, 
the last term with the exponential can be approximated by
$H^2 (t_1) / H^2 (t_k)$, and then~:
\begin{equation}
{\cal C}_{1} \simeq 
\frac{H(t_k)}{m_P^2 \sqrt{2 k^3}}
\frac{V(t_1)}{\frac{\partial V}{\partial A} (t_k)}.
\label{c1fora}
\end{equation} 
This approached result matches exactly the slow-roll prediction
(\ref{slowrollC}), because $V(t_1)$ is nothing else but $V_A$. In
fact, we find that the numerical result (which takes into account the
exact relation $- 2 m_P^2 \dot{H} =
\dot{\bar{A}}^2+\dot{\bar{B}}^2+\dot{\bar{C}}^2$, and the small
correction due to the fact that $\delta B_1$ cannot be neglected
during two $e$-folds) exceeds the slow-roll prediction by a factor~4.3.
  
To conclude on the evolution of $\Phi_1$, note that it is possible in
principle to isolate numerically the adiabatic and the isocurvature
contribution. We subtract to $\Phi_1$ a solution of the type of
eq.(\ref{exana}), and also, a solution $\tilde{\cal D}/a$. We tune
(${\cal C}_1, \tilde{\cal C}_1, \tilde{\cal D}$) in order to remove any
behavior proportional to $1/a$ or $H/a$, and to cancel $\Phi_1$
during the second slow-roll stage. The result is unique and stands for
the isocurvature mode.  The second plot of Fig. \ref{fig-} shows the
result of this decomposition.  At the end of this formal exercise,
$\Phi_1$ appears as the result of an almost exact cancellation during
the first stage between adiabatic and isocurvature contributions,
which are of opposite sign. We see that both modes are excited during
the transition, but the isocurvature mode decays like $a^{-3}$ during
the second stage and vanishes.

On the other hand, the solution for $j=2$ differs completely from the
standard slow-roll prediction. We expect from eqs.(\ref{slowrollC},
\ref{va}, \ref{vb}) that ${\cal C}_2 \sim 3 {\cal C}_1$. It is in fact
$10^3$ times smaller. Again, this result shows that the usual
definition of ($V_A$, $V_B$) is not valid during and before the
slow-roll interruption.  This appears clearly when we separate
numerically the adiabatic and isocurvature contribution (third plot of
fig.\ref{fig-}). We see that during the first stage, the isocurvature
mode completely dominates.

In conclusion, even if ${\cal C}=({\cal C}_1^2 + {\cal C}_2^2)^{1/2}$
differs from the slow-roll prediction only by a factor of order one,
we have proved that the usual results of eqs.(\ref{defvi},
\ref{slowrollC}) cannot be employed in our model. The numerical results can
be matched with eqs.(\ref{invertsys},~\ref{slowrollC}) by tuning $V_A$ and
$V_B$ during the first stage, but we don't know how to predict these
values analytically~: this calculation seems to be difficult, and
a numerical simulation is probably unavoidable.

\subsection{Small wavelength result}

We now consider the modes that cross the Hubble radius well after the transition.
For the smallest wavelengths, this happens during the last single-field slow-roll stage,
and the usual results of eqs.(\ref{slowrollC}, \ref{va}, \ref{vb})
are automatically valid. In our model, these modes correspond to very small
wavelengths today, not observable on cosmic scales.

We are more interested in modes crossing the Hubble radius during the
oscillatory period,\footnote{Since the oscillations of $B$ produce
oscillations in the effective mass of $C$, one may expect that $\delta
C$ modes could undergo a stage of parametric resonance, that would
show up in resonant metric amplification \cite{Bruce}.  This is not the
case because the field $C$ is very heavy, so that the relative
amplitude of the mass oscillations is kept very small.} when the
evolution of $H$ is driven by $\dot{\bar{B}}^2 \gg
\dot{\bar{A}}^2$. The evolution of $(\Phi_1, \Phi_2)$ for such a mode,
with Hubble crossing at $N \simeq -44$ (six $e$-folds after $N_c$), is
shown on the last plot of fig.\ref{fig-}.  Since we are now in
single-field inflation, $\Phi_1$ and $\Phi_2$ should be pure adiabatic
modes; we check that this is the case, by fitting ($\Phi_1$, $\Phi_2$)
with the expression of eq.(\ref{exana}) (for $N> -44$). The
coefficient ${\cal C}_1$ matches exactly the slow-roll analytic
prediction, and ${\cal C}_2 \ll {\cal C}_1$, as expected for a
non-slow-rolling field perturbation.

This can be explained easily.  During the transition, $\Phi_1$ remains
a slowly-varying solution, because the effective mass of $\delta A_1$
is dominated by $k^2/a^2$. So, at Hubble crossing, $\Phi_1$ is in the
slowly-varying adiabatic mode, and ${\cal C}_1$ is exactly the same as
in the slow-roll prediction (i.e. eq.(\ref{slowrollC}) with $i=1$).
On the other hand, $|\Phi_2|$ is strongly affected by the transition,
because the effective mass of $\delta B_2$ becomes suddenly much
larger than $k^2/a^2$. At Hubble crossing, $|\Phi_2|$ is in the
decaying adiabatic mode, proportional to $a^{-1}$, and stabilize with
a value much smaller than $|\Phi_1|$, as expected for a non
slow-rolling field perturbation in single-field inflation.

\subsection{Primordial power spectrum}

We plot on fig.\ref{figSP-} the power spectrum of adiabatic
fluctuations, $k^{3/2} {\cal C}(k)$, or equivalently (up to a
constant), $k^{3/2} \Phi(k)$.  It is a step-like spectrum with
superimposed oscillations on small scales, quite similar to the
analytic spectrum of Starobinsky \cite{S92}, but more smooth (the
amplitude of the first oscillations with respect to the amplitude of
the step is smaller). On intermediate and small scales, it is also
close to the spectrum of \cite{P92} (from double ``polynomial''
inflation), with the notable difference that on large scale we have an
approximately flat plateau, instead of a logarithmic dependence on
$k$.

For our choice of primordial parameters, the step has got an amplitude
$p \simeq 10$, and spans over one decade in $k$ space.  
As we saw before, this spectrum arises solely from
${\cal C}_1$, i.e., from metric perturbations calculated with initial
condition of the type ($\delta A \neq 0$, $\delta B=0$).

\section{Dynamics of the Higgs field}

Until now, we have treated the Higgs field $C$ as a simple classical
homogeneous background field, and put by hand an initial condition at
$N=N_c$, in order to push the field away from unstable equilibrium.
We will now enter into more details concerning the Higgs quantum
fluctuations, in order to justify and calculate the initial condition
at $N_c$ (by averaging over quantum fluctuations), and to prepare the
work of the following section, in which we will consider the effect of
the Higgs quantum perturbations on the primordial power spectrum.
We rewrite the potential (Appendix, eq.(\ref{potential})) under  the
more suggestive form~:
\begin{equation} \label{mexhat}
V = V(A,B) + \frac{1}{4} \beta^2 (B^2 - B_c^2) C^2 + 
\frac{\beta^2 B_c^2}{8 C_{eq}^2} C^4.
\end{equation}
When $B < B_c$, the potential is a usual ``Mexican hat'' with respect to $C$.

\subsection{Fluctuations during the first stage of inflation}

We decompose the field $C$ as in
eqs.(\ref{decomp},\ref{def.mode.fct}), using now a three-dimensional
operator basis $\hat{a}_{j,~j=1,2,3}$. Since $C$ is complex, we would
need in principle a four-dimensional basis, in order to quantify
separately the real and imaginary part $C_1$ and $C_2$. However, in
this section, the $U(1)$ symmetry will still be preserved, and the
mode functions are identical for $C_1$ and $C_2$. So, for concision,
we introduce only one degree of freedom $j=3$, describing
simultaneously both directions in the complex plane.

The Wronskian condition (\ref{wronskian}) applies also to the mode
function $\delta C_{j}$. During the first stage of inflation, 
$\bar{C}=0$. So, in addition to eqs.(\ref{eq.-}) (now taken for
$j=1,2,3$), we have the following perturbation equation~:
\begin{equation} \label{eq.mot.C}
\ddot{\delta C_{j}} + 3H \dot{\delta C_{j}} + 
\left( \frac{k^2}{a^2} + \frac{\partial^2 V}{\partial C^2} \right)
\delta C_{j} = 0,
\end{equation}
with a WKB solution valid both within
and outside the Hubble radius~:
\begin{eqnarray} 
\label{WKB2}
\delta C_{j} &=& \delta_{3j} \frac{a^{-3/2}}{\sqrt{2 m}} e^{-i \int m dt},
\\
{\rm with} \qquad
m^2 &\equiv& \frac{k^2}{a^2} + \frac{\partial^2 V}{\partial C^2} 
- \frac{9}{4} H^2. \nonumber
\end{eqnarray}
During the first stage of inflation, at the first perturbative order
considered here, $\Phi_{3}$ vanishes, and so do the non-diagonal mode
functions ($\delta A_3$, $\delta B_3$, 
$\delta C_{j \neq 3} $).  $\delta C_{3}$
is decoupled from other perturbations and its evolution depends mainly
on the mass $\partial^2 V / \partial C^2 =\frac{1}{2} \beta^2
(B^2-B_c^2)$ that goes to zero at $N=N_c$.

\subsection{Formation of inhomogeneities just after $N_c$} 

When $N>N_c$, the mass of $C$ becomes negative, causing an exponential
amplification of the mode functions, known as spinodal instability.
Let us restrict the analysis to the very short stage after $N_c$,
during which $\delta C_{3}$ remain small enough for the linear
equation (\ref{eq.mot.C}) to be still valid.\footnote{this is the case
if $\langle 0| \; | \int \frac{d^3 {\bf k}}{(2 \pi)^{3/2}} \delta C_k
|^2 |0 \rangle^{1/2} \ll C_{eq}$~: then, the cubic term in $\partial V
/ \partial C$ is negligible with respect to the linear term, and
$\frac{\partial V}{\partial C} ( \int \frac{d^3 {\bf k}}{(2
\pi)^{3/2}} \delta C_k ) \simeq \frac{\partial^2 V}{\partial C^2}
(0)~\times~\int \frac{d^3 {\bf k}}{(2 \pi)^{3/2}} \delta C_k$.} As we
can check {\it a posteriori}, this stage lasts $\sim 0.15$ $e$-folds.
If we re-scale $\delta C_{3}$ to $\chi \equiv a^{3/2} \delta C_{3}$,
we see that eq.  (\ref{eq.mot.C}) can be rewritten~:
\begin{equation} \label{chi}
\ddot{\chi} + \left( \frac{k^2}{a^2} + \frac{\partial^2 V}{\partial C^2}
- \frac{9}{4} H^2 \right)
\chi = 0.
\end{equation}
We will solve this equation under a few approximations, and compare our result with
an exact numerical simulation.
Since we are interested in a short period after $N_c$, with typically
$N-N_c \leq 0.2$, it is appropriate
to linearize the $N$-dependence of the
effective mass-term in equation (\ref{chi})~:
\begin{eqnarray}
\frac{\partial^2 V}{\partial C^2} &=& - \mu^2 (N-N_c), 
\qquad \mu^2 \equiv - 
\left. \frac{d \{ \partial^2 V / \partial C^2 \} }{dN} \right|_{N_c} > 0 , \\
\frac{k^2}{a^2} &=& \frac{k^2}{a_c^2}  (1-2(N-N_c)) \nonumber
\end{eqnarray}
(in the particular model we are studying, $\mu$ can be easily found
from the slow-roll condition for $\bar{B}$~: $ \mu^2 = \frac{\beta^2
B}{3 H^2} \frac{\partial V}{\partial B} = \frac{\beta^2}{3 H^2} \frac{
2 \ln2 (g_A^2\xi_A+g_B^2\xi_B)^2}{16 \pi^2} $).  Also, during this
short stage, $H$ is approximately constant and $N-N_c=H(t-t_c)$.
Under these approximations, equation (\ref{chi}) reads~:
\begin{equation} \label{chi2}
\ddot{\chi} - \mu_k^2 \; H \; (t-t_k) \;
\chi = 0,
\end{equation}
with~:
\begin{equation}
\mu_k^2 = \mu^2 + 2 \frac{k^2}{a_c^2},
\qquad
t_k = t_c + \frac{1}{\mu_k^2 H} \left( \frac{k^2}{a_c^2} - \frac{9}{4} H^2 \right).
\nonumber
\end{equation}
The modes $\chi$ start growing exponentially at $t=t_k$ (for $k=0$, 
$t=t_c - \frac{9 H}{4 \mu^2}$, but
smaller wavelengths start growing later).
The correctly normalized solution of equations (\ref{chi2}) and
(\ref{eq.mot.C}) is given, up to an
arbitrary complex phase, by~:
\begin{eqnarray} \label{bessel}
\chi &=& \sqrt{\frac{\pi}{6}} (t-t_k)^{1/2}~ 
{\cal H}^{(1)}_{1/3} \! \left[ \frac{2i}{3} \mu_k H^{1/2} (t-t_k)^{3/2} \right], \\
\delta C_{3} &=& a^{-3/2} \chi,
\nonumber
\end{eqnarray}
where ${\cal H}^{(1)}_{1/3} [z]$ is a Hankel function of the first kind.
Indeed, for $t < t_k$, this solution has an asymptotic expression
which is identical to equation (\ref{WKB2}). For $t \simeq t_k$, one has~:
\begin{equation} \label{bessel1}
\delta C_{3} = a^{-3/2} 
\frac{\sqrt{2 \pi}}{3 \Gamma (\frac{2}{3})}
\left( \frac{3}{\mu_k H^{1/2}} \right)^{1/3}, 
\end{equation}
and for $t > t_k$ in very good approximation~:
\begin{equation} \label{bessel2}
\delta C_{3} = a^{-3/2} 
\frac{\exp \left( \frac{2}{3} \mu_k H^{1/2} (t-t_k)^{3/2} \right)}
{\sqrt{2 \mu_k H^{1/2} (t-t_k)^{1/2}}}.
\end{equation}
Is it possible to define an effective background from the
coarse-graining of long wavelength modes? The field $C$, coarse-grained over
a patch of size $l \sim 2 \pi a / k$, is a stochastic
classical\footnote{The effective background can be considered as
classical when large wavelengths modes have very large expectation
values. Then, observers measuring the value and momentum of $C$ cannot
feel the non-commutative operator structure of the field, and the
quantum field behaves as a classical stochastic field.
More precisely, it is
know \cite{DECO} that the modes can be approximated by classical stochastic
quantities when their expectation values are much bigger 
than the minimal expectation value set by Heisenberg uncertainty principle~:
\begin{equation}
\left| \langle \Psi | \; \hat{y}({\bf k}) \hat{p}^{\dagger} ({\bf k}) 
+ {\rm c.c.} \; | \Psi \rangle \right|
\gg
\left| \langle \Psi | \; [ \hat{y}({\bf k}) , \hat{p}^{\dagger} ({\bf k}) ]
\; | \Psi \rangle \right| = 1.
\end{equation}
This condition on operators can be translated in terms of mode functions
and read for the vacuum state~:
\begin{equation} 
| \delta C \delta \dot{C}^*
+ \delta C^* \delta \dot{C} |
\gg -i( \delta C \delta \dot{C}^*
- \delta C^* \delta \dot{C}).
\end{equation}
For eq.(\ref{WKB2}) (before $N_c$) this means $H \gg m_C$, which
does not hold for the non slow-rolling field $C$; but when the
exponential amplification starts, the condition above is rapidly
satisfied for spinodal modes.}
quantity. Its real and imaginary part obey to a Gaussian distribution
(at least if all modes are in the vacuum state), with variance
$<\!\bar{C}^2\!>^{1/2}$ computed from~:
\begin{eqnarray} \label{cgi}
<\!\bar{C}^2 \!>
&\equiv& 
\int_{0}^{k_{cg}} \!\!\! \int 
\frac{d^3 {\bf k}~ d^3 {\bf k}'}{(2 \pi)^3} 
\langle 0 | 
\delta C ({\bf k}, {\bf x},t)
\delta C^{\dagger} ({\bf k'}, {\bf x},t)
| 0 \rangle
\nonumber \\
&=&  \frac{1}{2 \pi^2} \int_{0}^{k_{cg}} 
k^2 \left| \delta C_{3} \right|^2 dk,
\end{eqnarray}
where $k_{cg}$ is the coarse-graining cut-off.  When $t < t_c$, we
know that scales outside and around the Hubble radius have a
$k$-independent amplitude (see eq.(\ref{WKB2})). So, $<\!\bar{C}^2 \!>$
scales like $k_{cg}^3$ (intuitively, all modes have the same
amplitude, but small wavelength have a bigger statistical weight,
proportional to $k^2$). On the other hand, when $t \geq t_c$, large
wavelengths start being amplified earlier; so, in the coarse-graining
integral, there is a competition between the term $|\delta C_{3} |^2$
and the term $k^2$, and a scale will be privileged.  More precisely,
let us use the asymptotic expression (\ref{bessel2}). The function
$k^2 | \delta C_{3} |^2$ peaks around the scale\footnote{to
find this it is appropriate to use the approximation $k^2 | C_{3} |^2
\propto k^2 \exp(4/3 \; \mu \; H^{1/2} (t-t_c-\frac{k^2}{\mu^2 a_c^2}
)^{3/2} )$.}
\begin{equation}
k_{max}= a_c \left( \frac{\mu H}{2 \sqrt{N-N_c}} \right)^{1/2},
\label{kmax}
\end{equation}
and only modes in the range $0.1 k_{max} < k < 3 k_{max}$, which are
within the Hubble radius, contribute to the coarse-graining integral
(\ref{cgi}). A numerical simulation confirms this result for $0.1 \leq
N-N_c \leq 0.2$ (see fig.\ref{figCG}).  
This means that just after $N_c$, the field $C$ has
an inhomogeneous structure, and can be seen as effectively homogeneous only
in regions of typical size $\lambda \ll a / k_{max}$.

Inside each patch, we can define an effective
homogeneous background from the coarse-graining of large wavelength
modes, and use the standard semi-classical approximation. 
The constant phase in a given patch can be chosen randomly, while the
squared modulus $|\bar{C}|$ obeys a probability distribution
that can be calculated exactly.
Indeed, with any coarse-graining cut-off $k_{cg} \sim 3 k_{max}$, we
can compute the root mean square of the real and imaginary part of
$C$~:
\begin{equation} 
<\!\bar{C}^2 \!>^{1/2} = \frac{1}{4 \pi^{3/2}} \left( \frac{a_c}{a} \right)^{3/2} 
\left( \frac{\mu_0^2 H}{t-t_c} \right)^{1/4}
\exp \left( \frac{2}{3} \mu H^{1/2} (t-t_c)^{3/2} \right).
\label{likezm}
\end{equation} 
Since $C_1$ and $C_2$ are Gaussian stochastic numbers with variance $<
\! \bar{C}^2 \! >^{1/2}$, then $|C|^2$ obeys a $\chi^2$ distribution
and $|C|$ has a mean value $\simeq \sqrt{3/2}
<\!\bar{C}^2\!>^{1/2}$ and a variance $\simeq \sqrt{1/2}
<\!\bar{C}^2\!>^{1/2}$ (the numerical factors can be found in
tables; see, for instance, eq.(26.4.34) in \cite{AS}).  We see explicitly, by comparing with
eqs.(\ref{bessel1}, \ref{bessel2}), that $<\!\bar{C}^2\!>^{1/2}$ is a
solution of the zero-mode equation with initial condition~:
$<\!\bar{C}^2 \!>^{1/2} (t_c) = (6 \pi \Gamma (\frac{2}{3}))^{-1}
(3 \mu^2 H)^{1/3}$. Actually, the numerical simulation shows that the
exact result exceeds this estimate by a factor 2.3, so that the
correct initial condition is~:
\begin{equation} 
\label{iniC}
<\!\bar{C}^2 \!>^{1/2} (t_c) = 0.12~(\mu^2 H)^{1/3}.
\end{equation}
So, we really have in each patch an effective homogeneous background,
obeying for $N \geq 0.1$ to the equation $\ddot{\bar{C}} + 3H
\dot{\bar{C}} + \partial V / \partial C =0$, and with an initial
$\chi^2$ distribution of probability.
%Note that in
%contrast with the case of ``spinodal inflation'', in which the Higgs
%is in slow-roll, the effective zero-mode arises mainly from the averaging
%of sub-Hubble scales.

On the other hand, on scales $k \leq k_{max}$, the background is essentially 
inhomogeneous, and due to strong mode-mode coupling, the linear semi-classical 
approximation breaks soon after $N_c$.

\subsection{Later evolution and the homogeneous phase approximation}

So, it is not possible to find an exact solution for the evolution of
$C$ during spinodal instability and at later time.  We should recall
here that out-of-equilibrium phase transitions and spinodal
instability have been widely studied in condensed matter physics, and
also in the context of ``spinodal inflation'', using either the
large-N limit \cite{BVH1,BVH2} (which is well-defined, but appropriate
for the symmetry breaking of $U(N)$, $N \rightarrow \infty$), or just
a Hartree-Fock approximation \cite{CH} (which is more difficult to
interpret, and cannot account so far for the metric back-reaction). In
such studies, the typical scale of homogeneity for the coarse-grained
background is larger than the Hubble radius ($k_{max} < a(t_c)
H(t_c)$), while the observable primordial spectrum arises from modes
with $k > a(t_c) H(t_c)$. So, it is possible to separate completely
the modes contributing to the ``effective zero-mode assembly'', and
those contributing to the primordial spectrum. In contrast, in our
case, we need to follow modes on scales $k$ with $k < a(t_c)H(t_c) <
k_{max}$. So, at first sight, it seems that our problem is much more
complicated than spinodal inflation, due to the inhomogeneity of the
background. In fact, it is not, because we have an inflaton field $A$
driving inflation even during the transition, and we will see that a
perturbative semi-classical approach can still be
employed.\footnote{Another difference with respect to spinodal
inflation is that we will never coarse-grain or average modes
far outside the Hubble radius. This is just because the spinodal field
is not slow-rolling, and then, modes are exponentially amplified also on
sub-Hubble scales.}

The following observations will suggest an approximation under which we can 
continue the primordial spectrum calculation~:
\begin{itemize}
\item
since the causal horizon during the first inflationary stage is much
bigger than the Hubble radius, ($A$, $B$) are homogeneous on the
largest scales observable today, $B_c$ is reached everywhere at the
same time, and the phase transition is triggered coherently. So, the
modulus $|C|$ should be approximately homogeneous, even on the largest
scales~: we should be able to cast it into a zero-mode plus small
perturbations.
\item 
the phase inhomogeneities should not be much relevant for the
primordial spectrum calculation. The existence of a negative mass,
leading to exponential amplification, only arise in the longitudinal
direction $|C|$. The degree of freedom associated with the complex
phase is a Goldstone boson; it is eaten up by the gauge field, which
becomes massive. Starting from zero, this mass soon exceeds the Hubble
parameter~: so, there cannot be large quantum fluctuations arising from
this sector. The existence of phase inhomogeneities is also linked to
the formation of cosmic strings at the very beginning of spontaneous
symmetry breaking, separated at least by a characteristic length
$\lambda > a/k_{max}$. These strings could eventually contribute to
the formation of cosmological perturbations today, but as we said in the
introduction, we do not include such mechanism in this study.
\end{itemize}

So, for our purpose, it seems reasonable to neglect the phase
inhomogeneities, forget the vacuum degeneracy, and assume a
homogeneous phase $\theta({\bf x},t) = \bar{\theta}$. In the following
section, we will recompute the primordial power spectrum, taking now
into account the Higgs longitudinal quantum perturbations, and we will
find that these perturbations have a characteristic signature on the
primordial power spectrum.

\section{Spectrum with Higgs longitudinal perturbations}

We go on neglecting the vacuum degeneracy. We saw in section 4.2 that
just after the transition, the modulus $|C|$ has got a mean value
$\sqrt{3/2} <\!|C|^2\!>^{1/2}$ and a variance $\sqrt{1/2}
<\!|C|^2\!>^{1/2}$ that we calculated explicitly. Assuming a
homogeneous phase is almost equivalent to identifying this modulus
with a real scalar field, canonically quantized, and decomposed,
following the usual semi-classical approximation, into a background
field $\bar{C}= \sqrt{3/2} <\!|C|^2\!>^{1/2}$, plus quantum
fluctuations. The quantum mode functions should be matched with the
quantities $\delta C_j$ that we already studied around $N_c$, divided
by $\sqrt{2}$. With these prescriptions, the mean value and the
variance of the modulus previously studied in the last section
coincide exactly with those of the real scalar field introduced in
this paragraph.\footnote{at least, when $(N-N_c)>0.1$ (i.e., when
$<|C|^2>^{1/2}$ obeys the zero-mode equation, see eq.(\ref{likezm})),
and $(N-N_c)<0.2$ (i.e., when the equation for $\delta C_j$ remains
linear, see eq.(\ref{eq.mot.C})).}

This matching is the key point of our study. It is not exact, because,
in a classical stochastic sense, it is based only on the first two
momenta of the $\chi^2$ distribution~: we artificially ``gaussianize''
the fluctuations. However, the matching seems appropriate for
estimating the amplitude of the quantum fluctuations of the modulus
$|C|$, even if any information on a possible non-gaussianity is lost.
Since in this approach we introduce a (physically justified)
zero-mode, we expect that the growth of spinodal modes will slow-down
and terminate rapidly, and that the perturbative semi-classical
approach will remain valid. This will be justified explicitly in
section 5.3.

In summary, we simulate the usual background equations for 
($\bar{A}$, $\bar{B}$, $\bar{C}$, $H$) and the following perturbation 
equations~:
\begin{eqnarray}
\ddot{\delta A_j} + 3H \dot{\delta A_j} + 
\left( \frac{k^2}{a^2} + \frac{\partial^2 V}{\partial A^2} \right)
\delta A_j
&=& 4 \dot{\bar{A}} \dot{\Phi}_j - 2 \frac{\partial V}{\partial A}
\Phi_j - \frac{\partial^2 V}{\partial A \partial B}
\delta B_j, \nonumber \\
\ddot{\delta B_j} + 3H \dot{\delta B_j} + 
\left( \frac{k^2}{a^2} + \frac{\partial^2 V}{\partial B^2} \right)
\delta B_j
&=& 4 \dot{\bar{B}} \dot{\Phi}_j - 2 \frac{\partial V}{\partial B}
\Phi_j
- \frac{\partial^2 V}{\partial A \partial B} \delta A_j
- \frac{\partial^2 V}{\partial B \partial C} \delta C_j, 
\nonumber \\
\ddot{\delta C_j} + 3H \dot{\delta C_j} + 
\left( \frac{k^2}{a^2} + \frac{\partial^2 V}{\partial C^2} \right)
\delta C_j
&=& 4 \dot{\bar{C}} \dot{\Phi}_j - 2 \frac{\partial V}{\partial C}
\Phi_j
- \frac{\partial^2 V}{\partial B \partial C} \delta B_j,
\nonumber \\
\dot{\Phi}_j + H \Phi_j &=& \frac{1}{2 m_P^2} (
\dot{\bar{A}} \delta A_j + \dot{\bar{B}} \delta B_j 
+ \dot{\bar{C}} \delta C_j).
\end{eqnarray}
When we start following a mode $\delta C_j$ well inside the Hubble
radius, we employ the initial condition of eq.(\ref{WKB2}) divided by
a factor $\sqrt{2}$. At $N=N_c$, we put by hand a non zero-value for
$\bar{C}$, given by eq.(\ref{iniC}) multiplied by a factor
$\sqrt{3/2}$.

\subsection{Large wavelength results}

The evolution of ($|\Phi_1|$, $|\Phi_2|$, $|\Phi_3|$) for a mode
crossing the Hubble radius at $N \simeq -53$ (i.e. during the
oscillatory stage) is shown on the upper plot of fig.\ref{fig+}.  By
subtracting the previous solution (fig.\ref{fig-}) to the new solution
(fig.\ref{fig+}), we check that the evolution of $|\Phi_1|$ is
unchanged, except during a few $e$-folds after the beginning of the
transition. The difference is due to the small (one-loop) coupling
term $\frac{\partial^2 V}{\partial A \partial B}$~: when $\delta B_1$
is amplified, it excites $|\Phi_1|$, but later on, this mode decays
and doesn't leave an observable signature.  On the other hand, the
negative effective mass of $\delta C_j$ and the large (tree-level)
coupling between $\delta B_j$ and $\delta C_j$ generate a considerable
amplification of $|\Phi_2|$ and $|\Phi_3|$. At the end, when the
isocurvature modes and the decaying adiabatic mode can be neglected,
the remaining growing adiabatic mode is smaller than the one for
$|\Phi_1|$, but only by one order of magnitude.

So, on large scale, what is the observable difference between these
results, and those of section 3 ?  First, since initial conditions for
$\delta B_2$ are proportional to $(a \sqrt{k})^{-1}$, we expect the
large-scale power spectrum for $k^{3/2} |\Phi_2|$ to be, as usual,
approximately scale-invariant. In other words, the contribution found
for one mode will be the same for other large-scale modes. So, the
large-scale power spectrum will still be a flat plateau, but possibly
with a small deviation from gaussianity, caused by the small
contribution of $|\Phi_2|$. As we said before, our method does not
allow a quantitative evaluation of this deviation, and we leave this
for further studies. Second, since initial conditions for $\delta C_3$
are independent of the scale, and since the factor $k^2/a^2$ is always
sub-dominant in the effective mass of $\delta C_3$ on large
wavelengths, $|\Phi_3|$ should be scale-independent on these
wavelengths. So, the contribution of $k^{3/2} |\Phi_3|$ to the power
spectrum $k^{3/2} <|\Phi|^2>^{1/2}$ will be proportional to $k^{3/2}$,
and will peak for scales crossing the Hubble length during the
transition.  We therefore expect an observable non-Gaussian spike,
whose shape and amplitude will be calculated in section 5.4.

\subsection{Small wavelength result}

The evolution of ($|\Phi_1|$, $|\Phi_2|$, $|\Phi_3|$), for a mode
crossing the Hubble radius at $N \simeq -43$ (i.e. during the
oscillatory stage), is shown on the lower plot of fig.\ref{fig+}.
$|\Phi_1|$ and $|\Phi_2|$ are identical to previous results of section
3.2.  $|\Phi_3|$ has the same behavior as $|\Phi_2|$, but is even
smaller. So, for small wavelengths, including the Higgs modulus
perturbation makes no difference at all.

\subsection{Consistency check of the semi-classical approximation}

Before giving the primordial spectrum of perturbations, we will check
here the consistency of our approach, and the validity of the semi-classical
approximation in our case. Strictly speaking, the semi-classical
approximation holds whenever for each field, the Klein-Gordon equation 
\begin{equation}
g^{-1/2} \partial_{\mu} (g^{1/2} g^{\mu \nu} \partial_{\nu}
\varphi ( {\bf x}, t)) + 
\frac{\partial V}{\partial \varphi} \left( \varphi ( {\bf x}, t) \right)=0
\end{equation}
can be casted into a background equation for the zero-mode, plus
independent linear equations for field and metric mode functions.  The
first term, with space-time derivatives, is linear provided that
metric perturbations remain small~:
\begin{equation}
<\!0|~| \int \frac{d^3 {\bf k}}{(2 \pi)^{3/2}} \Phi |^2 |0\!>^{1/2}
\ll 1.
\end{equation}
The second term, with the potential derivative, is linear in two cases~:
\begin{enumerate}
\item 
obviously, when
the perturbations are small with respect to
the zero-mode~: \\
$<\!0|~| \int \frac{d^3 {\bf k}}{(2 \pi)^{3/2}} \delta \varphi 
|^2 |0\!>^{1/2} \ll \bar{\varphi} $.
Then, we can write~:
\begin{equation}
\frac{\partial V}{\partial \varphi} \left( \varphi( {\bf x}, t) \right)
\simeq
\frac{\partial V}{\partial \varphi} (\bar{\varphi}) \; + \;
\frac{\partial^2 V}{\partial \varphi^2} (\bar{\varphi}) ~\times~
\int \frac{d^3 {\bf k}}{(2 \pi)^{3/2}} \delta \varphi.
\end{equation}
\item
also, when both the zero-mode and the perturbation expectation value
are small enough for non-quadratic terms in the potential to be neglected.
Then, we have the following linearization~:
\begin{equation}
\frac{\partial V}{\partial \varphi} \left( \varphi( {\bf x}, t) \right)
= \frac{\partial^2 V}{\partial \varphi^2} \times \varphi( {\bf x}, t),
\end{equation}
and $\partial^2 V / \partial \varphi^2$ is a constant.
\end{enumerate}

We have seen that around $N=N_c + 0.15$, in the homogeneous phase
approximation, we can define a zero-mode and a set of
perturbations. Since all these quantities are very small, we are in
the second case, and the semi-classical approximation is valid. But
what happens later? We know from figure \ref{fig+} that one $e$-fold
after $N_c$, the perturbations of $C$, $B$ and $\Phi$ reach a maximum;
if, at that time, the expectation values of the perturbations exceed
the homogeneous background fields (or, in the case of metric
perturbation, exceed one), then our approach is simply inconsistent.

To address this question, we first study the evolution of the power
spectra of $\delta C_j$, $\delta B_j$ and $\Phi_j$, for $N_c < N < N_c
+ 1$. The results for $\delta C_j$ are shown on three-dimensional
plots, in fig.\ref{fig3D}.  This simulation is just the continuation
of the one performed in section 4.2, and at low $N$ we recognize the
curves of fig.\ref{figCG}. On the upper plot, we show $\sum_j <|\delta
C_j|^2>$, in logarithmic scale, vs. $k$ and $N$. We see that large
scales (spinodal modes) are exponentially amplified during the
spinodal stage, and then perform coherent oscillations, driven by the
background evolution.  On the lower plot, we show $k^2 \sum_j <|\delta
C_j|^2>$, in linear scale, vs. $k$ and $N$. We observe that the main
contribution to the averaging integral (eq.(\ref{cgi})) always arise
from the same region in $k$ space, centered around the approximately
constant wavenumber $k_{max}$. This observation is technically
important, because it justifies the use of finite limits of
integration in the averaging integral. In particular, we see that we
don't have to worry with ultra-violet divergences.  Provided that the
upper limit of integration $k_{cut-of\!f}$ is chosen in a finite range
(namely, around the largest $k$ values shown of fig.\ref{fig3D}), then
$<|C|^2>$ does not depend on this cut-off. Similar results are found
for $\delta B_j$ and $\Phi_j$.

Then, by integrating on $k$, we find the expectation values
$<|C|^2>^{1/2}$, $<|B|^2>^{1/2}$, $<|\Phi|^2>^{1/2}$, and plot them
with respect to $\bar{C}$, $\bar{B}$ and $1$ (see fig.\ref{figCCH}).
It appears that for $N \simeq N_c+0.20$, when $\bar{C}$ approaches the
valley of minima (on the figure, we recognize this valley from the
small oscillations around it), i.e., when $\partial V / \partial C$
becomes non-linear, the quantum perturbations are sub-dominant by one
order of magnitude. When $\bar{C}$ reaches $C_{eq}$, the quantum
expectation value is smaller by three orders of magnitude. After, we
know from fig.\ref{fig+} that the field will become more and more
homogeneous.  For $B$, we see that the perturbations are always
sub-dominant (when $\bar{B}$ performs its first oscillation, it goes
through zero, but the perturbations are still quite small and we
remain in the linear regime). For $\Phi$, the maximal value is
$<|\Phi|^2>^{1/2} \sim 10^{-3}$. So, the semi-classical approach
holds at any time in the range $N_c < N < N_c+1$, and therefore, until
the end of inflation.

%Let us conclude this subsection with a general remark. The validity of
%the semi-classical approximation could be suspected (at least for $C$)
%from the moment at which we introduced a zero-mode, of the same order
%of magnitude as the perturbation expectation value. Then, we knew that
%during a short time, the evolution of $\bar{C}$ and $<|C|^2>^{1/2}$ would
%be comparable, and that afterwards, the zero-mode would push the field
%out of the spinodal region, putting an end to the exponential amplification
%of the modes, and making the field $C$ more and more homogeneous.

In conclusion, in this model, the semi-classical approximation can be
employed during the whole evolution, because there exists a time ($N
\simeq N_c+0.15$) at which the perturbations have grown sufficiently
to create an effective zero-mode (obeying the usual zero-mode
equation), but not too much, so that we are still in the linear region
of the potential derivative. It is at this time\footnote{this time can
be clearly identified on the upper plot of fig.\ref{figCCH}~: without
the factor $\sqrt{3/2}$ (resp. $\sqrt{1/2}$) in $\bar{C}$
(resp. $<|C|^2>^{1/2}$), the curves would be tangent around $N=N_c +
0.15$.} that we have done the matching between the complex field (with
no zero-mode) and a real scalar field (with a zero-mode) standing for
its modulus. If such a time did not exist, complicated non-linear
approaches and further approximations would have to be employed. This
would be the case if the second inflaton field $A$ was not supplying a
constant potential energy, and supporting the Universe
expansion. Indeed, in that case, the evolution of the scale-factor and
of the fields would become completely stochastic, and it would be a
hopeless challenge to keep track of modes crossing the Hubble length
before and during the transition.

\subsection{Primordial spectrum including Higgs longitudinal perturbations}

We show our final result on figure \ref{figSP+}, for the parameters
used previously (middle).  On the other plots, we show some
primordial spectra corresponding to different values of $\beta$. The
contribution of $|\Phi_1|$ is unchanged with respect to
fig. \ref{figSP-}, but we now have an additional contribution from
$\Phi_2$ and $\Phi_3$ on intermediate and large scales.  Let us
summarize the characteristics of the power spectrum~:

$\bullet$
On small scales (large $k$), we have an approximately flat plateau,
exactly identical to the analytic predictions. Its amplitude can be
found in our previous paper (\cite{DD}, eqs. (6,7)), as a function of
the Fayet-Illiopoulos terms ($\xi_A$, $\xi_B$), of the coupling
constants ($g_A$, $g_B$) and of the duration of the second stage
$\Delta N$ (different values of $\Delta N$ correspond to different
choices for the inflaton fields initial values). The plateau has got a
tilt $n_S = 1 - 1 / \Delta N \simeq 0.98$, like in usual
(single-field) inflationary models, when the potential derivative is
provided by logarithmic one-loop corrections.

$\bullet$ On intermediate scales (the value of these scales is given
by $\Delta N$), we find a spike, as expected from sections 5.1 and 5.2
(since $k^{3/2} |\Phi_3|$ is proportional to $k^{3/2}$ for small $k$,
and strongly suppressed for large $k$). This spike may be
non-Gaussian, for reasons that we already explained. Clearly, it is
beyond the scope of this paper to quantify the non-gaussianity,
because the semi-classical approximation is not appropriate for
extracting the statistics of the modes.  The amplitude and width of
this spike with respect to large scales depend essentially on
$\beta$. Indeed, $\beta$ is proportional to the Higgs mass, and then,
controls the duration of the transition~: for low $\beta$ values, the
transition is long, exponential amplification is enhanced, and the
spike is higher. Its width is roughly of one decade in $k$ space (but,
since $k^{3/2} |\Phi_3|$ is proportional to $k^{3/2}$ on large scales,
small spikes are automatically more narrow). The shape of the spike
around its maximum is slightly non-trivial, reflecting the first
oscillations of the field $B$ around zero during the transition. It is
reasonable to state that if the primordial power spectrum really
possesses a spike (in the range in which this is still an open
possibility, {\it i.e.}, $0.1 \times 10^{-2} h$Mpc$^{-1} \leq k \leq 5
\times 10^{-2} h$Mpc$^{-1}$), then its amplitude is already
constrained by observations to be smaller than $\sim$10 (here, we
define the amplitude as the ratio between the maximum value of
$k^{3/2} |\Phi|$ and the one, for instance, of the large-scale
plateau).  This corresponds to the condition $\beta \geq 10^{-3}$.

$\bullet$ On large scales (small $k$), we also have an approximately
flat plateau, but as we saw in section 3.1, we don't know how to
predict analytically the amplitude and tilt of this plateau, due to
the violent slow-roll interruption during the transition. For our
choice of parameters ($\xi_A$, $\xi_B$, $g_A$, $g_B$, $\Delta N$),
there is a ratio of 4.3 between the amplitude obtained numerically for
$k^{3/2} |\Phi|$, and the analytic prediction (eq.(13) of \cite{DD},
or eq.(\ref{c1fora}) of this paper). This ratio is found to be
approximately independent of $\beta$. The large scale tilt is very
close to one. Finally, metric perturbations on these scales are
composed of a (Gaussian) component $\Phi_1$, and of a (possibly
non-Gaussian) component $\Phi_2$, generated, like $\Phi_3$, by the
exponential amplification of perturbations during the transition.  So,
the relative contribution of $\Phi_2$ depends, like the spike, on
$\beta$.  For $\beta \geq 10^{-3}$, we find that $\Phi_2$ is
sub-dominant, as can be seen in the figures, but not negligible (for
$\beta=10^{-3}$, $|\Phi_1| / |\Phi_2|=3.4$).  We also investigate the
dependence of the large-scale plateau on $\xi_A$, $\xi_B$. Since
small-scale amplitude depends on $\xi_A-\xi_B$ (see eq.(7) of
\cite{DD}), we vary $\xi_A+\xi_B$ while keeping $\xi_A-\xi_B$
fixed. We find that the amplitude on large and intermediate scales
increases with $\xi_A+\xi_B$. On fig.\ref{figSP++}, $\xi_A+\xi_B$ has
been divided by 2.8, and other parameters are the same as in
fig.\ref{figSP+} (upper plot).  In this case, the ratio between the
two plateaus is 1.5.

\section{Conclusion}

We calculated numerically the primordial power spectrum for a
particular model of double supersymmetric inflation, called double
D-term inflation \cite{DD}.  Following the usual semi-classical
approach, we integrated for each independent mode function the
equations of motion of field and scalar metric perturbations.  Some
technical difficulties (attached to any model of double hybrid
inflation) arise between the two inflationary stages. Indeed, the
first stage ends with a rapid second-order phase transition, taking
place during inflation, and corresponding to the spontaneous symmetry
breaking of a $U(1)$ gauge symmetry, triggered by a Higgs field
$C$. This transition is followed by another inflationary stage,
equivalent to usual hybrid inflation.  We were especially interested
in the behavior of modes exiting the Hubble radius during the
transition, since they are expected to produce non-trivial features in
the primordial power spectrum.

After spontaneous symmetry breaking, it is impossible to follow
exactly the mode functions of the complex field $C$.  Indeed, on the
largest observable wavelengths, the complex phase of $C$ is completely
inhomogeneous, at least at the beginning of the phase transition; so,
on these scales, the usual perturbative semi-classical approach is
broken by mode-mode coupling.  However, we argued that only
longitudinal Higgs perturbations, {\it i.e.}, perturbations of $|C|$,
can contribute to the primordial power spectrum. Indeed, these
perturbations undergo a stage of spinodal instability, during which
large wavelength modes are exponentially amplified, while the
perturbations of the complex phase and of the gauge field (which
combine into a massive gauge field, following the usual Higgs
mechanism), are those of a non-slow-rolling field with large positive
mass square.

So, we performed a calculation in which longitudinal Higgs
perturbations are taken into account, with the modulus $|C|$ treated
as an ordinary real scalar field, described by the semi-classical
equations. We showed that at the beginning of the transition, a
zero-mode for $|C|$ emerges from quantum fluctuations of $C$'s real and
imaginary parts.  Since the final result depends crucially on this
initial zero-mode, we calculated it precisely in section 4.2. Then, by
comparing the background fields and the averaged quantum
perturbations, we showed that the semi-classical approach holds at any
time. However, during one $e$-fold or so after the beginning of the
transition, it is not far from its limits of validity (with typically
a factor $10^{3}$ between background quantities and
perturbations). This fact is a first hint that the contribution to the
power spectrum resulting from longitudinal Higgs perturbations could
be significantly non-Gaussian. The second hints comes from the fact
that at the beginning of the transition, we identified a modulus with a
real scalar field. In this process, we matched carefully the mean
value and expectation value of both quantities, but information on
higher momenta was lost. It is beyond the scope of this paper to
evaluate the non-gaussianity.

Our final power spectrum, shown on fig.\ref{figSP+} for different
parameter values, can be divided in three regions~:
\begin{enumerate}
\item 
a small-scale Gaussian
plateau, whose amplitude and tilt ($n_S \simeq 0.98$) can be calculated
analytically.
\item 
a possibly non-Gaussian spike, with an amplitude depending on the
superpotential coupling parameter $\beta$. For $\beta \sim 10^{-3}$,
the spike induces variations at most of order one in the primordial
spectrum.
\item
a large-scale plateau, for which we don't have an exact analytic prediction.
For $\beta \sim 10^{-3}$, deviations from gaussianity, if any, are expected to 
be small.
\end{enumerate}
For a precise comparison of our model with CMB and
large-scale-structure (LSS) data, one needs in principle, not only the
primordial power spectrum of scalar adiabatic perturbations, but also
the spectrum of primordial gravitational waves, and the contribution
of cosmic strings formed during the two phase transitions (the one
considered here, and the one occurring at the end of
inflation). However, as we said in \cite{DD}, the tensor-to-scalar
ratio $T/S$ (for the first temperature anisotropy multipoles) depends
essentially on the gauge coupling constants (as in single-field D-term
inflation). Both $g_A$ and $g_B$ should be at most of order $10^{-1}$
(if not, inflation takes place with inflaton values of the order of
the Planck mass, and supersymmetry/supergravity theories are not
expected to be valid; see the Appendix for the consistency of this
requirement with string theories). With this order of magnitude for
the coupling constants, one can show easily that $T/S$ is negligible
in double D-term inflation. On the other hand, cosmic strings can
contribute significantly to the matter power spectrum and to CMB
anisotropies, especially for D-term inflation, as shown in \cite{RJ}
(see also \cite{CHM}). Since, in current studies, perturbations from
strings and from the primordial spectrum are just added in quadrature,
it is perfectly reasonable, at least in a first step, to calculate
only the primordial spectrum. The ratio of strings-to-inflation
large-scale temperature anisotropies, $R_{SI}=C_5^S / C_5^I$, is
difficult to estimate; theoretical predictions give a ratio around 3
\cite{RJ} or 4 \cite{CHM}, but the authors of this last reference
argue that there are many uncertainties in these calculations, so that
smaller values can still be considered seriously.
%(if $R_{SI}$ was bigger
%than one, D-term inflation would presumably be ruled out by current
%evidence for a first CMB peak, and by constraints on the matter power
%spectrum, at least in the framework of standard CDM). 
With a mixture of inflationary and string perturbations, our model
could still be interesting~: the features would appear nevertheless,
but smoothed by the cosmic string contribution.

In this work, we did not compare our results with CMB and LSS
observations. Indeed, current data cannot distinguish with precision
the kind of small features that we obtain, at least if they are
located on scales $0.1 \times 10^{-2} h$Mpc$^{-1} \leq k \leq 5 \times
10^{-2} h$Mpc$^{-1}$. However, as we said in the introduction,
comparisons with observations of typical BSI spectra (with a step or a
spike) have already been performed. The conclusions are currently
quite encouraging (see for instance \cite{LPS1,SG}) and provide strong
motivations for studying BSI inflationary models, until a final
statement is made by future redshift surveys, such as the 2-degree
Field project\footnote{http://meteor.anu.edu.au/colless/2dF} or the
Sloan Digital Sky
Survey\footnote{http://www.astro.princeton.edu/BBOOK}, combined with
forthcoming balloon and satellite CMB experiments.

An important qualitative aspect of our results is the possible
deviation from gaussianity on intermediate and large scales. This may
be related with the evidence for a non-gaussianity in the COBE data
\cite{CHM}, but further work is needed on this point, since, on the
one hand, we don't have a precise theoretical prediction of
non-gaussianity in this model, and on the other hand, the conclusion
of \cite{CHM} involves large uncertainties.

Finally, we should stress that experimental evidence for a BSI
primordial power spectrum would be a very positive breakthrough for
cosmology. On the one hand, the introduction of one or two additional
inflationary parameters, associated with the shape of the spike or the
step, would not compromise cosmological parameter extraction, as
indicated by \cite{LPP}, because the effect of these parameters would
be orthogonal to the other's. On the other hand, experimental data
would encode more information, and provide additional exciting
constraints on inflationary models, fundamental parameters, and
high-energy physics.

\section*{Acknowledgements}

This work was initiated after illuminating discussions with David
Polarski and Alexei Starobinsky. I would also like to thank Bruce
Bassett, Rachel Jeannerot, Stephane Lavignac and Andrei Linde for
providing useful comments, and Antonio Masiero for stimulating
discussions. This work is supported by the European Community under
TMR network contract No. FMRX-CT96-0090.

\newpage

\section*{Appendix}

{\bf The model.}  Double D-term inflation \cite{DD} is
based on the superpotential $ W= \alpha \tilde{A} \tilde{A}_+
\tilde{A}_- + \beta \tilde{B} \tilde{B}_+ \tilde{B}_- $, and on the
definition of two gauge groups ($U(1)_A$, $U(1)_B$), with
corresponding gauge coupling constants ($g_A$, $g_B$) and
Fayet-Iliopoulos terms ($\xi_A$, $\xi_B$). We choose the charges to be
(0,0) for $\tilde{A}$, $\tilde{B}$, ($\pm 1$, 0) for
$\tilde{A}_{\pm}$, and ($\pm 1$,$\pm 1$) for $\tilde{B}_{\pm}$ (other
choices can also lead to successful models).  During
inflation, we need only to follow the fields $\tilde{A}$, $\tilde{B}$
and $\tilde{B}_-$.  Moreover, the complex phases of $\tilde{A}$ and
$\tilde{B}$ remain constant during inflation (except for very peculiar
initial conditions that we don't consider here).  So, we deal
with four canonically normalized real fields ($A$, $B$, $C_1$,
$C_2$) defined as~:
\begin{equation}
A \equiv \sqrt{2} |\tilde{A}|, 
\qquad 
B \equiv \sqrt{2} |\tilde{B}|, 
\qquad 
C_1 + i C_2  \equiv C \equiv \sqrt{2} \tilde{B}_-.
\end{equation}
The scalar potential reads~:
\begin{eqnarray}
\label{potential}
V &=& \frac{g_A^2}{2} \left( \xi_A - \frac{1}{2} C^2 \right)^2
+ \frac{1}{4} \beta^2 B^2 C^2 
+ \frac{g_B^2}{2} \left( \xi_B - \frac{1}{2} C^2 \right)^2
\nonumber \\
&+& \frac{1}{64 \pi^2} \sum_{i=1...9} 
a_i m_i^4 \ln \left(\frac{m_i^2}{\Lambda^2}
\right).
\end{eqnarray}
The last term is the one-loop correction. Following \cite{JL},
($a_i$, $m_i$) are taken from the following table~:
\begin{center}
\begin{tabular}{|c|c|l|}
\hline
$i$ & $a_i$ & $m_i^2$ \\
\hline
1 & 2 & $\frac{1}{2} \alpha^2 A^2 + g_A^2 (\xi_A- \frac{1}{2} \langle C \rangle^2)$ \\
\hline
2 & 2 & $\frac{1}{2} \alpha^2 A^2 - g_A^2 (\xi_A- \frac{1}{2} \langle C \rangle^2)$ \\
\hline
3 & 2 & $\frac{1}{2} \beta^2 \langle C \rangle^2$ \\
\hline
4 & 2 & $\frac{1}{2} \beta^2 (2 B^2 + \langle C \rangle^2)$ \\
\hline
5$^{(>)}$ & 2 & $\frac{1}{2} \beta^2 B^2 - g_A^2 \xi_A - g_B^2 \xi_B$ \\
~$^{(<)}$ & 1 & $2 (g_A^2 \xi_A + g_B^2 \xi_B - \frac{1}{2} \beta^2 B^2)$ \\
\hline
6$^{(>)}$ & 2 & $0$ \\ 
~$^{(<)}$ & 3 & $(g_A^2 + g_B^2) \langle C \rangle^2$ \\
\hline
7 & -4 & $\frac{1}{2} \alpha^2 A^2$ \\
\hline
8 & -4 & $\left( \frac{1}{4} \beta^2 B^2 + \frac{1}{4} (\beta^2 + 2 g_A^2 + 2 g_B^2)
\langle C \rangle^2 + \sqrt{\Delta} \right)$ \\
\hline
9 & -4 & $\left( \frac{1}{4} \beta^2 B^2 + \frac{1}{4} (\beta^2 + 2 g_A^2 + 2 g_B^2)
\langle C \rangle^2 - \sqrt{\Delta} \right)$ \\
\hline
\end{tabular}
\end{center}
Here, $^{(>)}$ (resp. $^{(<)}$) means ``when $|B| \geq B_c$''
(resp. ``when $|B| \leq B_c$''), and $B_c = \sqrt{\frac{2(g_A^2 \xi_A
+ g_B^2 \xi_B)}{\beta^2}}$. In the table, we have used the notations~:
\begin{eqnarray}
\langle C \rangle &\equiv & \sqrt{
\frac{2(g_A^2 \xi_A + g_B^2 \xi_B - \frac{1}{2} \beta^2 B^2)}{(g_A^2 + g_B^2) }}, 
\label{ellipse} \\
\Delta &\equiv & \frac{1}{16} \left( \beta^2 B^2 + (\beta^2 + 2 g_A^2 + 2 g_B^2)
\langle C \rangle^2 \right)^2 - \frac{1}{2} \beta^2 (g_A^2 + g_B^2) 
\langle C \rangle^4. 
\end{eqnarray}
The one-loop corrections are automatically continuous in $|B| = B_c$.
The renormalization scale $\Lambda$ must be chosen around $\Lambda
\sim \beta B_c$ and, as in \cite{JL}, we optimize this choice numerically
by requiring the continuity of the potential derivative.

{\bf Choice of parameters.} The dependence of the primordial spectrum
on the parameters is discussed in section 5.4. Before this section, we
choose some arbitrary parameters $(\xi_A, \xi_B, g_A, g_B)$ inside
the allowed region defined in \cite{DD}, for which the power spectrum
has got the same order of magnitude as the one indicated by COBE, and
the amplitude of the step (between the large-scale and the small-scale
plateau) is of order one~:
\begin{equation}
\sqrt{\xi_A}= 3 \times 10^{-3} m_P, \qquad
\sqrt{\xi_B}= 4.2 \times 10^{-3} m_P, \qquad
g_A = g_B = 10^{-2}.
\end{equation}
The value of $\alpha$ is completely irrelevant for our study (it is
relevant only at the very end of inflation). On the other hand, the
choice of $\beta$ is crucial. Until section 5.3, we take
$\beta= 10^{-3}$. As mentioned in section 3, footnote 6, values of
$\beta$ greater than $10^{-2}$ would render the problem very
complicated; we would not have a robust justification for the
expression of one-loop corrections, and the transition would consist
in chaotic oscillations in ($B$, $C$) space.

{\bf String motivations}.  Double D-term inflation is a consistent
model from the point of view of supersymmetry, when the
Fayet-Iliopoulos terms are put by hand from the beginning. Here, we
briefly address the issue of consistency with string theory. It turns
out that in heterotic string theory, one can always redefine the gauge
groups in order to end up with only one anomalous $U(1)$, and a single
associated Fayet-Iliopoulos term. So, in this framework, it is
impossible to obtain two $\xi$ terms at low energy. However, it is
well-known that even the simplest models of single D-term inflation
are hardly compatible with heterotic string theory \cite{LyR,ERS},
mainly because the Fayet-Iliopoulos term generated by the
Green-Schwartz mechanism is typically too large to account for COBE
data. It has been noted by Halyo \cite{H} that the situation is much
better in type I string orbifolds, or type IIB orientifolds. Then, the
Fayet-Iliopoulos terms are not generated at the one-loop order
\cite{P}, but at tree level, and depend on the vacuum expectation
value of some moduli \cite{IRU} corresponding to the ``blowing-up
modes'' of the underlying orbifold; since there is some freedom in
fixing the moduli, the value of $\xi$ required by COBE does not appear
as unnatural in this framework (also, in these theories, there is no
unification of all coupling constants as in heterotic string theory,
and one can envisage scenarios where the coupling constant $g$ of the
anomalous gauge symmetries is slightly lowered with respect to the
Standard Model coupling constant \cite{H,Ibanez}, so that the
inflaton takes values significantly smaller than the Planck mass
during inflation, and supergravity, or even global supersymmetry can
be employed). But, nicely, another unusual feature in type I string
orbifolds and orientifolds of type IIB is the coexistence of several
anomalous $U(1)$'s, with associated Fayet-Iliopoulos terms
\cite{IRU}. So, our model may naturally arise from these theories. We
are very grateful to Stephane Lavignac for having kindly provided
information on these points.

\begin{figure}[p]
$$
\epsfxsize=13cm
\epsfbox{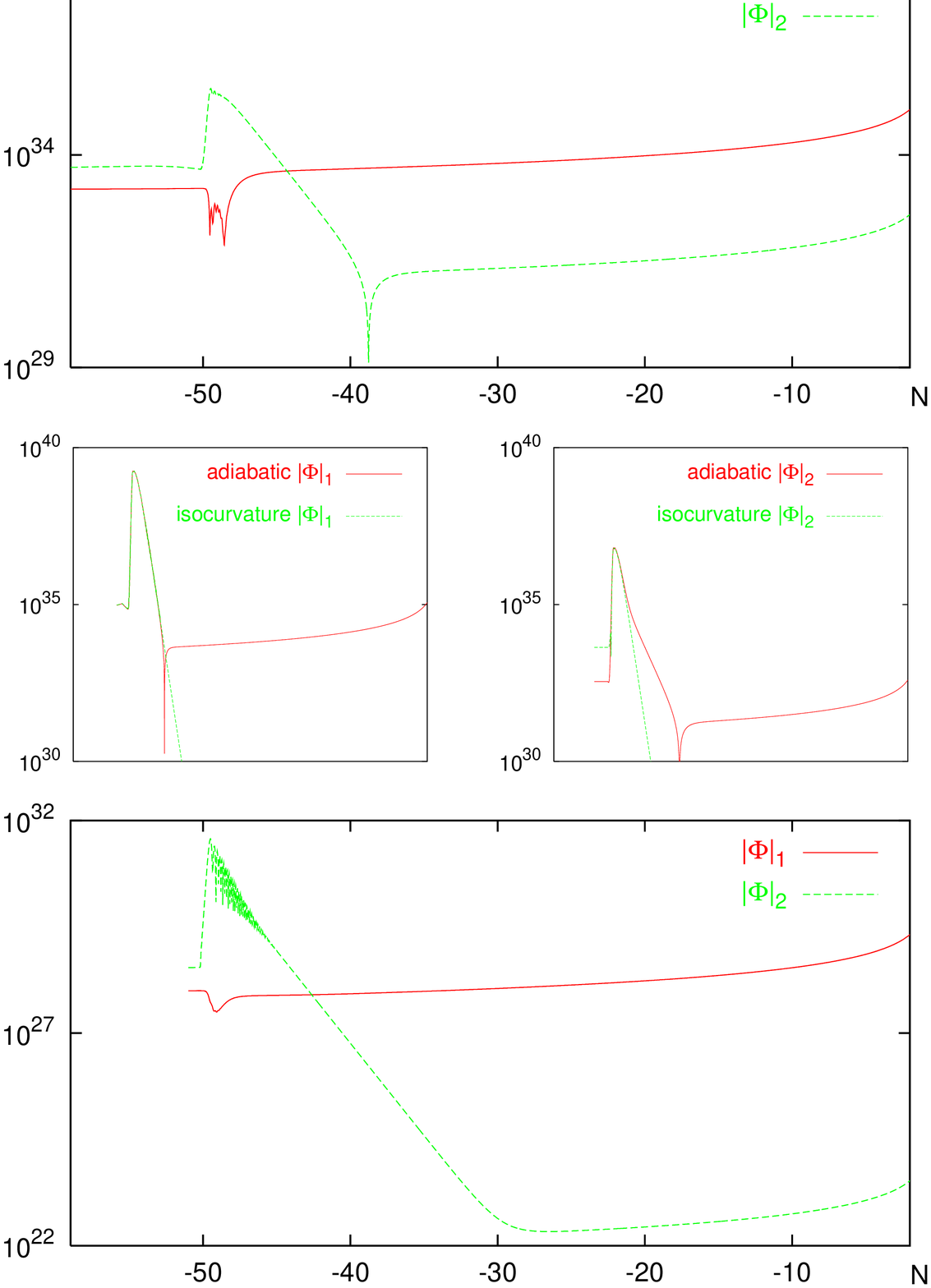}
$$
\caption[]{Evolution of the metric perturbations, when Higgs perturbations are
neglected~:
(top) $|\Phi_1|$ and $|\Phi_2|$ for a long wavelength mode;
(middle left) splitting of $|\Phi_1|$ in adiabatic and isocurvature contributions
for this mode;
(middle right) same for $|\Phi_2|$;
(bottom) $|\Phi_1|$ and $|\Phi_2|$ for a small wavelength mode.
The transition starts at $N_c \simeq -50$, while the last inflationary
stage ends at $N \simeq 0$. The long wavelength mode exits the Hubble radius
at $N=-53$, the small wavelength mode at $N=-44$.}
\label{fig-}
\end{figure}

\begin{figure}[p]
$$
\epsfxsize=13cm \epsfbox{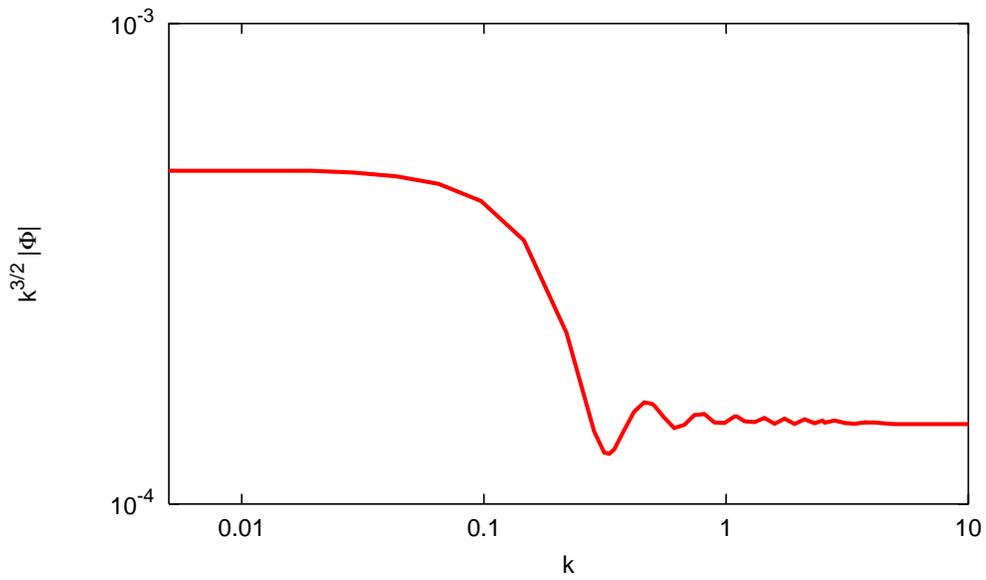}
$$
\caption[]{The primordial power spectrum of adiabatic fluctuations,
$k^{3/2} {\cal C}(k)$, or equivalently, $k^{3/2} | \Phi(k) |$, when
the Higgs perturbations are neglected (with logarithmic scale and arbitrary units).
It is a step-like spectrum with superimposed oscillations on intermediate scales.}
\label{figSP-}
\end{figure}

\begin{figure}[p]
%\vspace{-4cm}
%\mbox{ }
%\vspace{-0.8cm}
$$
\epsfxsize=6cm \epsfbox{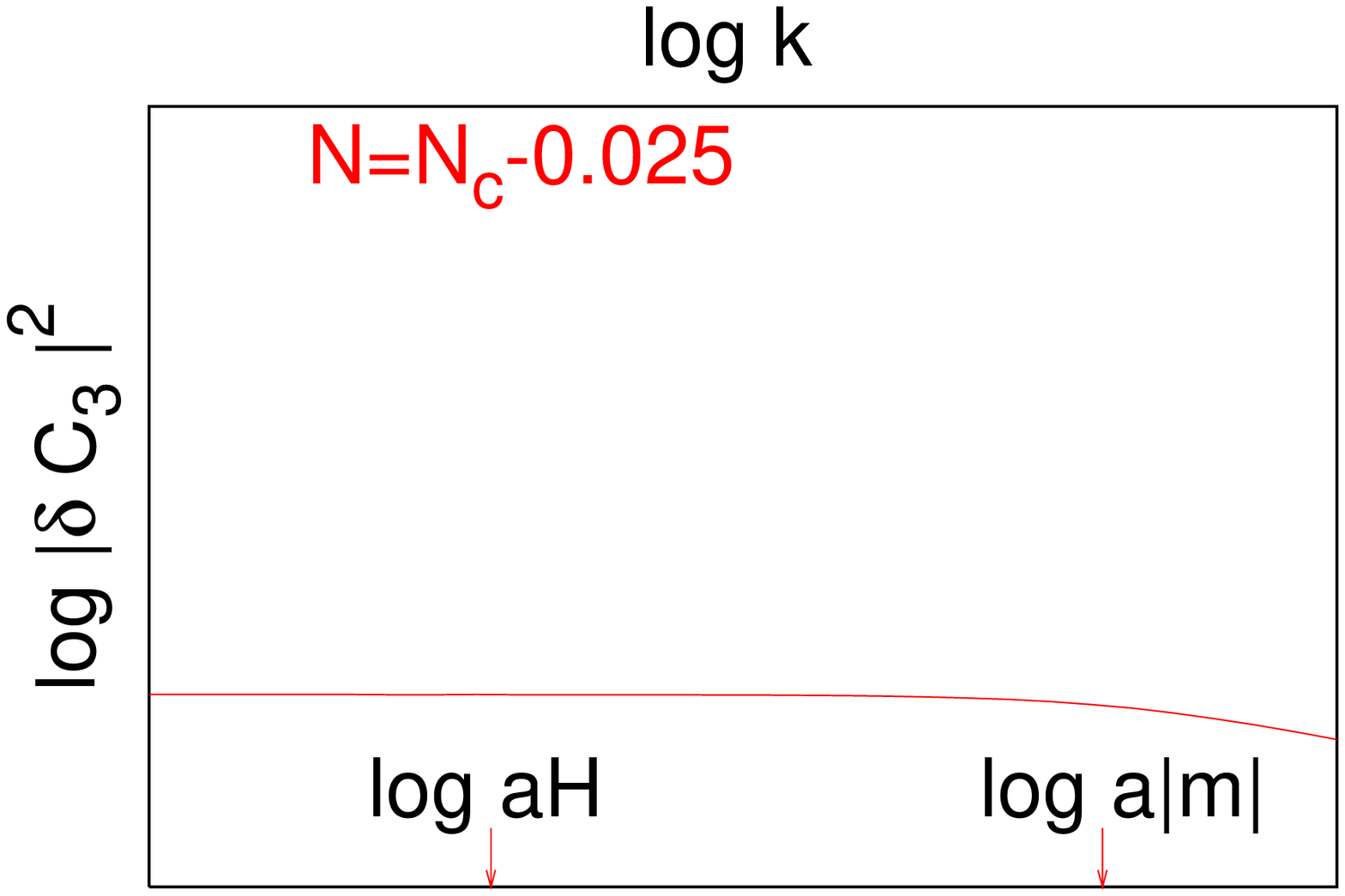}
\epsfxsize=6cm \epsfbox{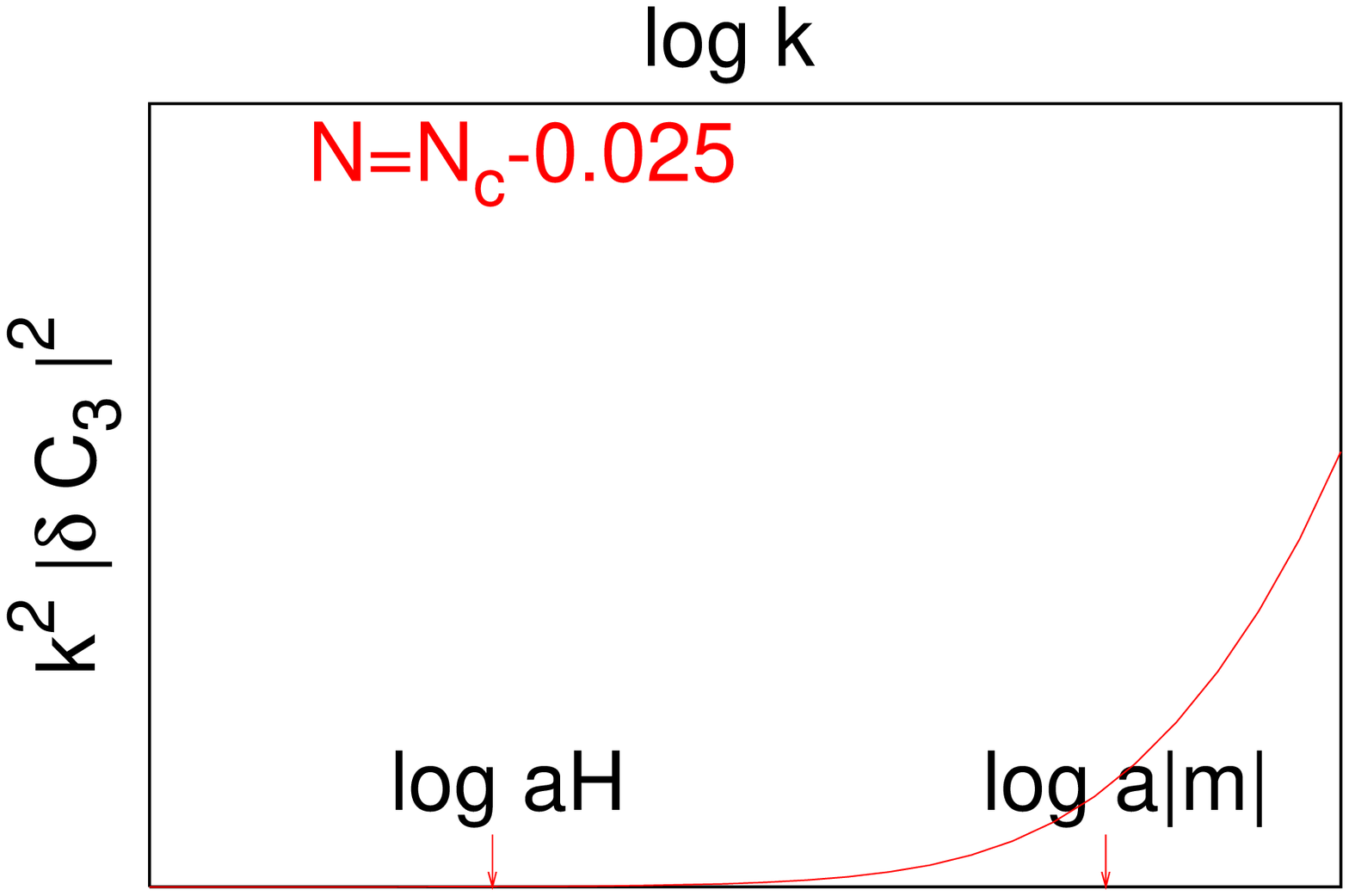}
$$
\vspace{-1.5cm}
$$
\epsfxsize=6cm \epsfbox{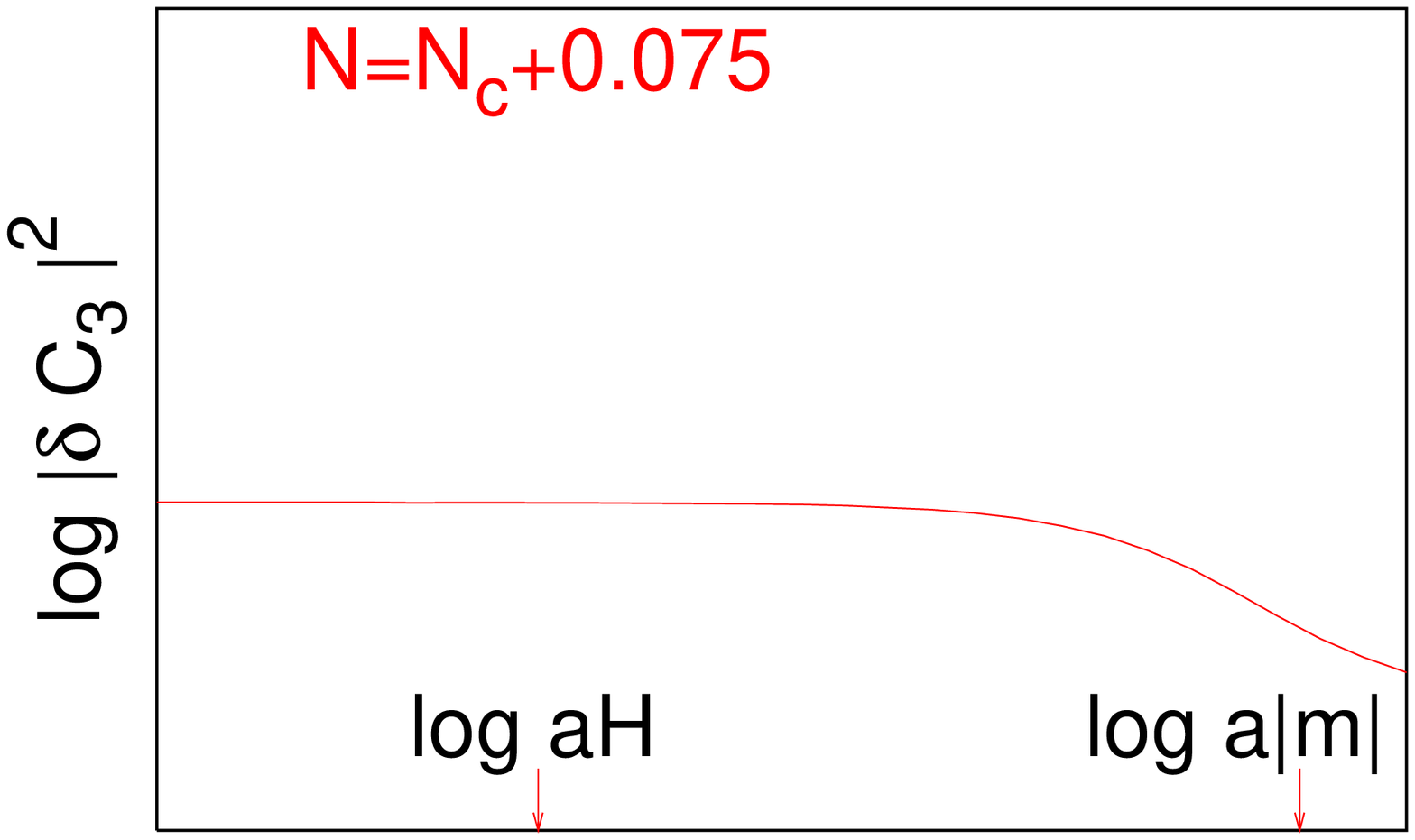}
\epsfxsize=6cm \epsfbox{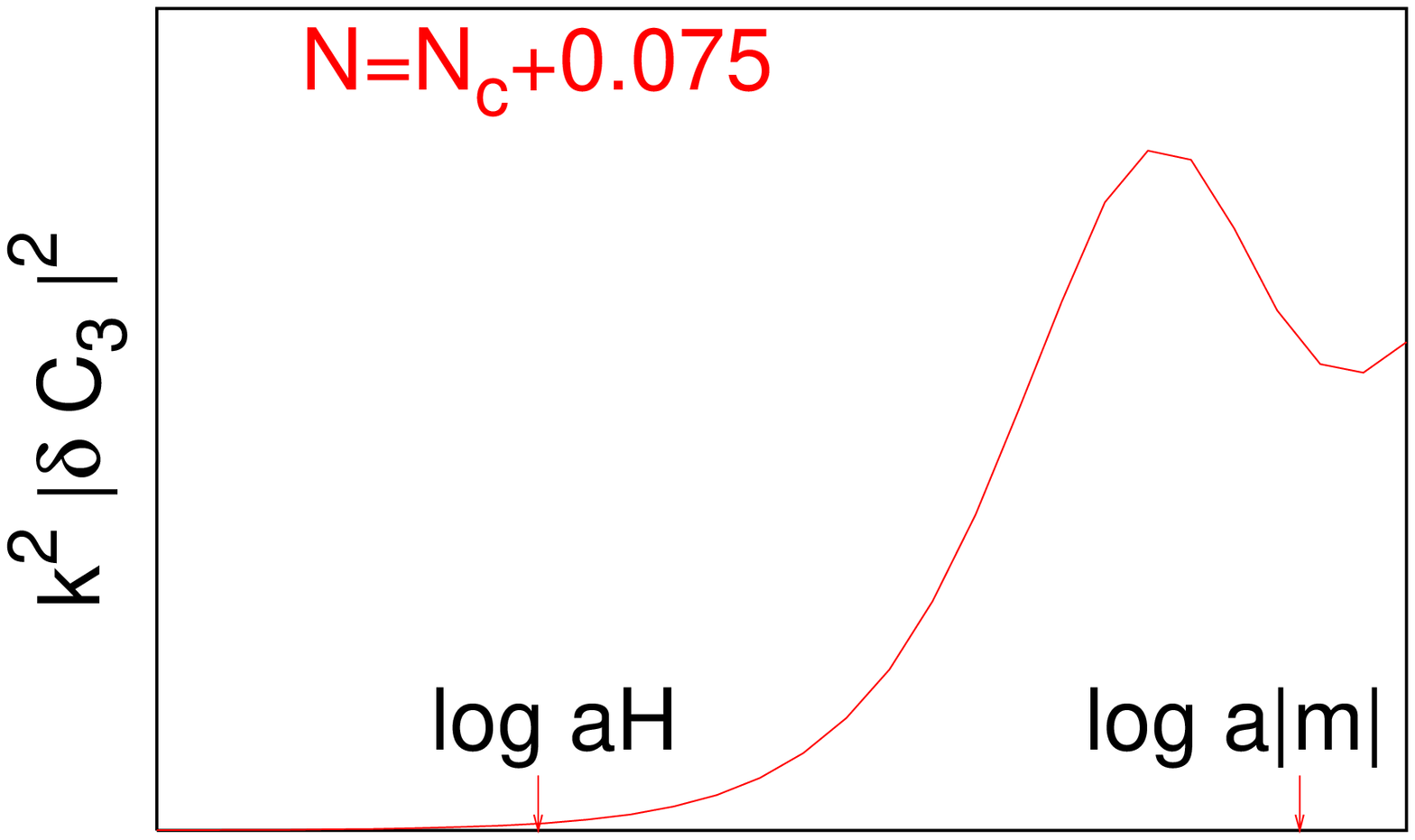}
$$
\vspace{-1.5cm}
$$
\epsfxsize=6cm \epsfbox{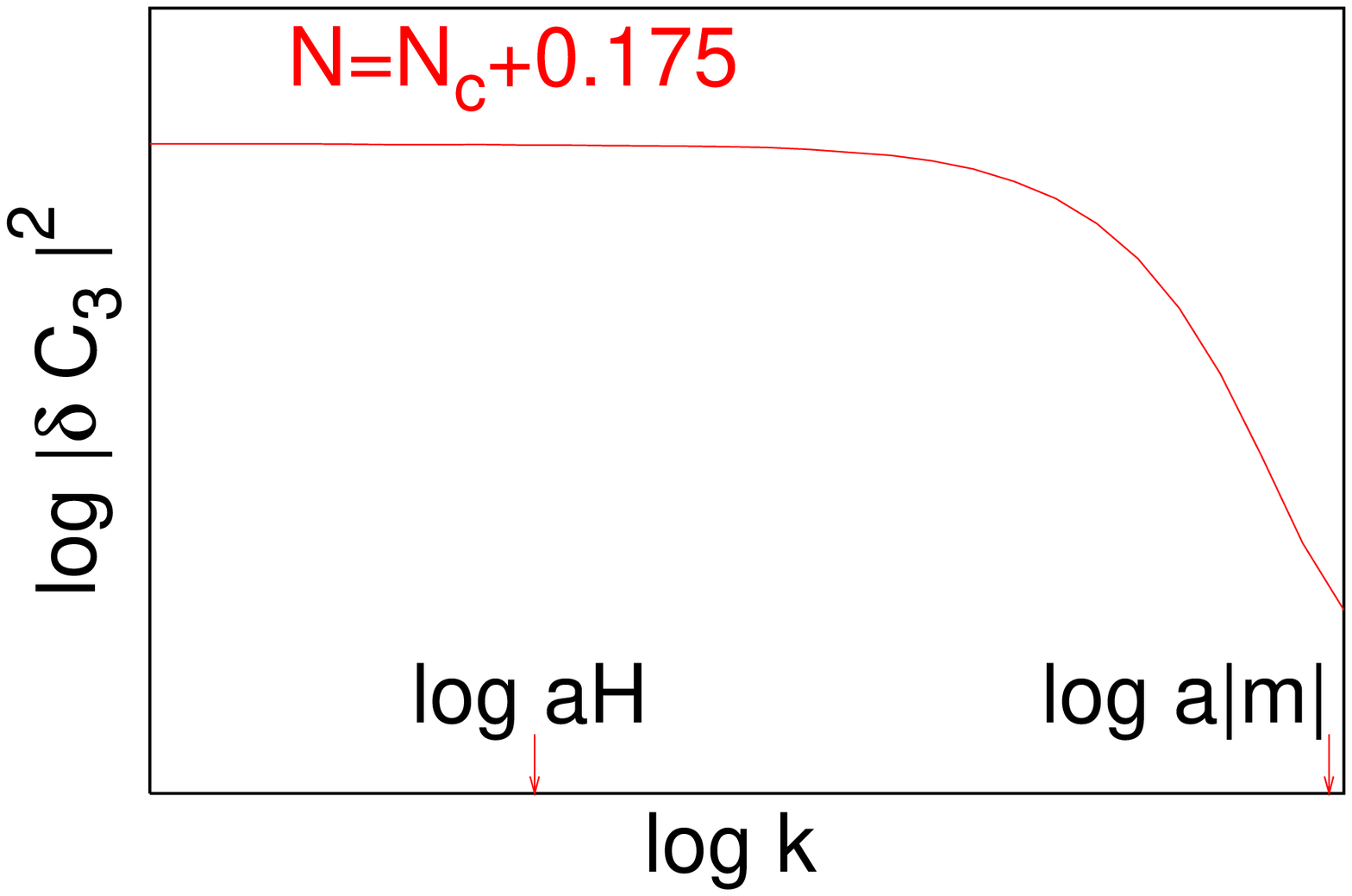}
\epsfxsize=6cm \epsfbox{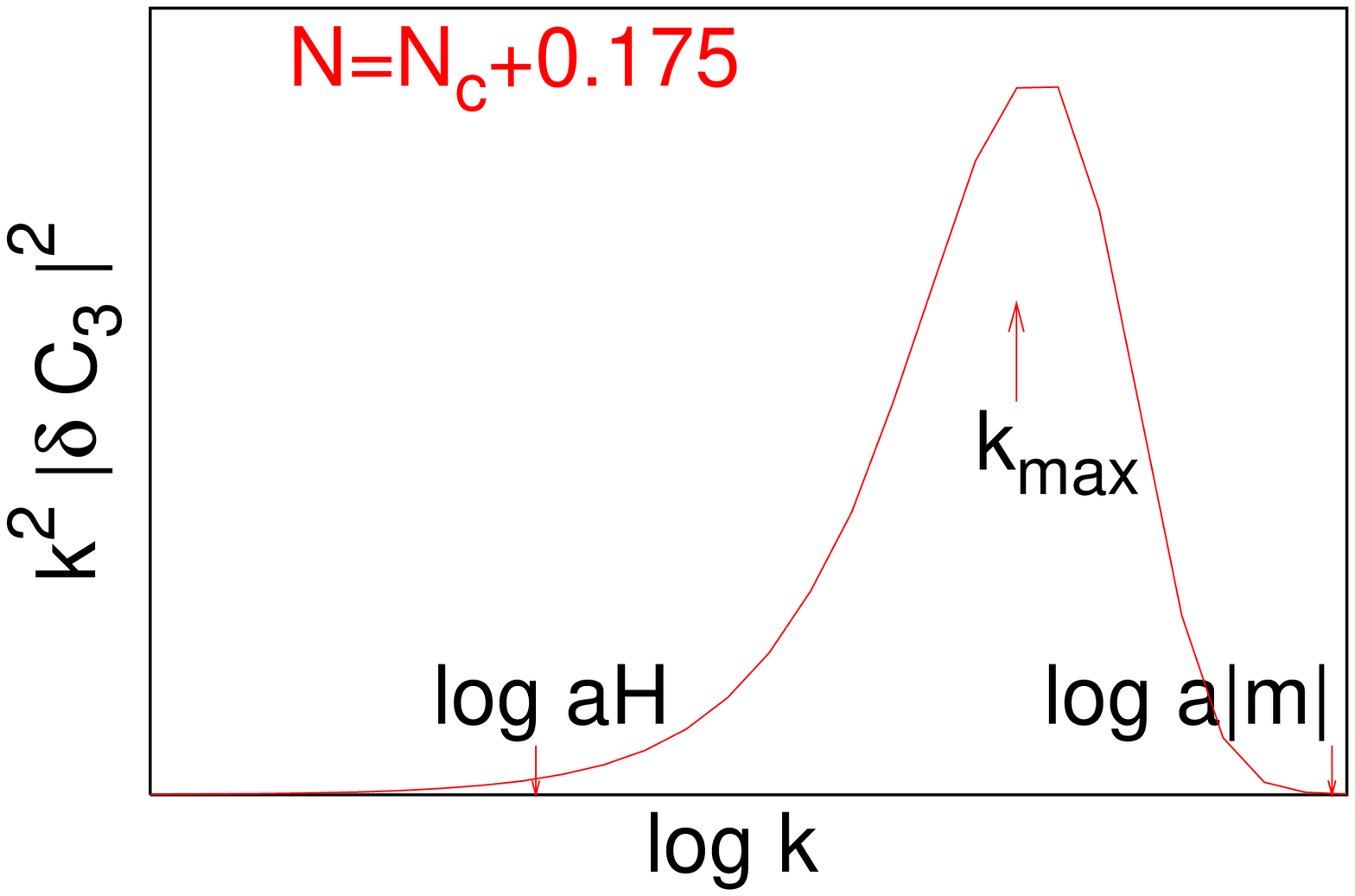} \nonumber
$$
\caption[]{The left column shows
the Higgs perturbation power spectrum~: $\log |\delta C_3|^2$ versus $\log k$,
for different times chosen around the beginning of the transition, i.e, 
when the effective mass of $C$ becomes negative. From top to bottom, we
have $N-N_c=-0.025, 0.075, 0.175$. The vertical scale is the same
in the three plots. We show on the horizontal axis the particular values
corresponding to $k=aH$ and $k=a |m|$, where $m^2\equiv \partial^2 V / \partial C^2$.
We see that spinodal modes, with $k \leq a |m|$, are amplified.

~~~On the right column, we plot the integrand of the coarse-graining
integral~: $k^2 |\delta C_3|^2$. The vertical scale increases from top
to bottom.  It appears clearly that after a fraction of $e$-fold, only
modes with $H < k/a < |m|$ contribute to this integral, which indicates
the emergence of an inhomogeneous background, with first-order
homogeneity recovered on smaller scales (higher $k$).  On the last
plot, we also show the scale $k_{max}$ calculated analytically from
eq.(\ref{kmax}).}
\label{figCG}
\end{figure}

\begin{figure}[p]
%\vspace{-4cm}
$$
\epsfxsize=13cm \epsfbox{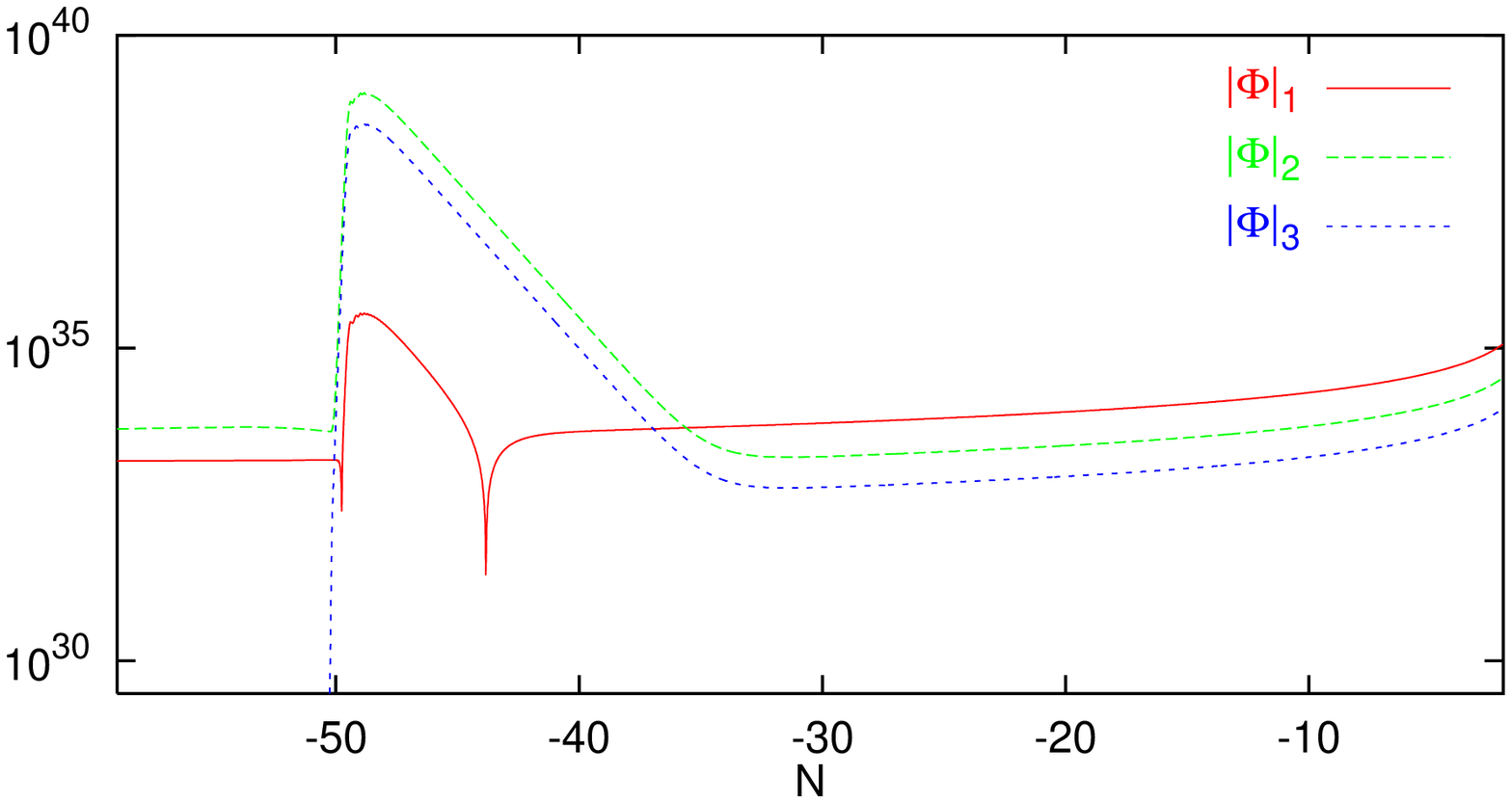}
$$
$$
\epsfxsize=13cm \epsfbox{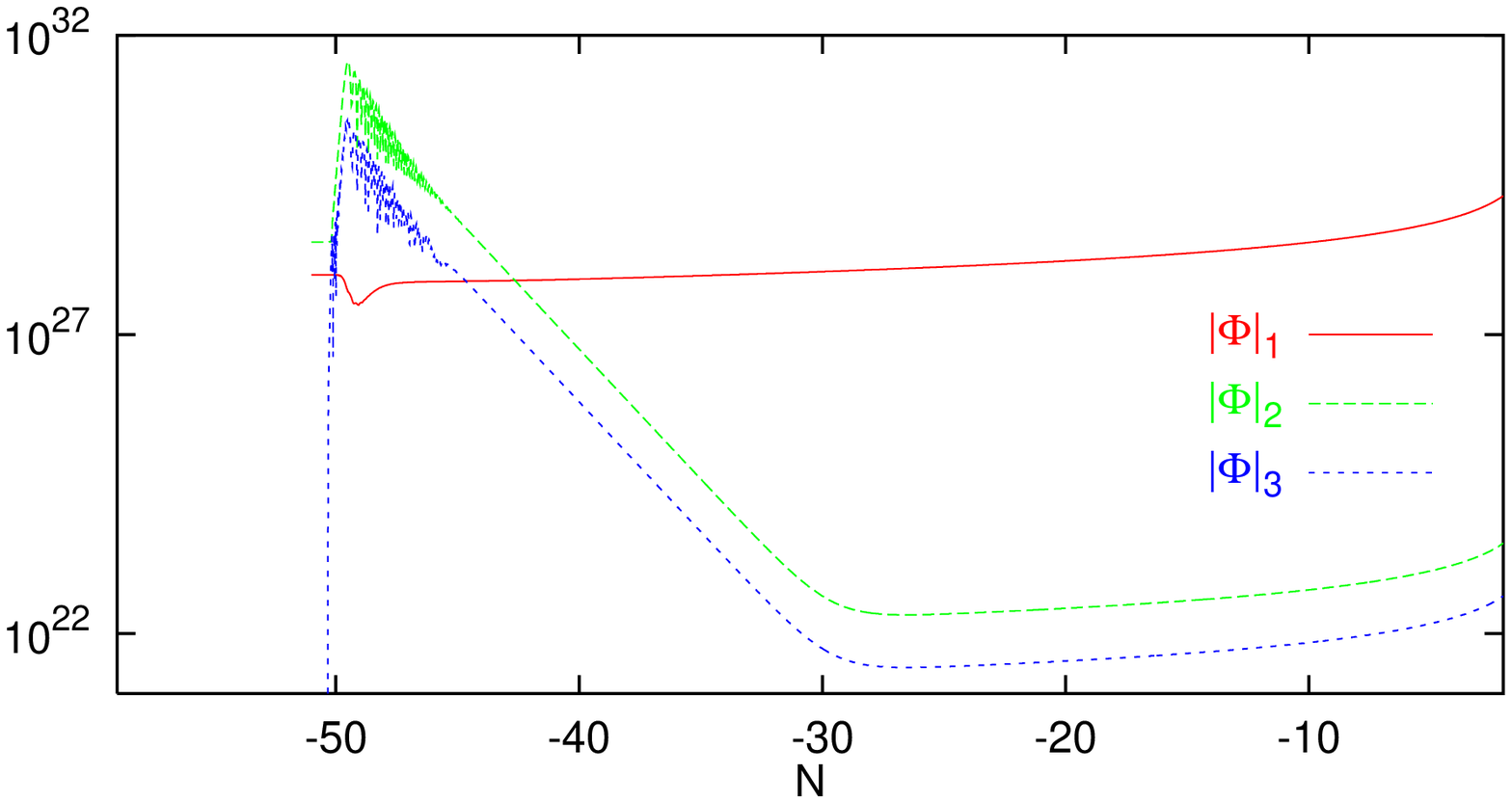} 
$$
%\nonumber 
%\\
%\epsfxsize=6cm \epsfbox{SIMULS/modes59-2.ps}&
%&\epsfxsize=6cm \epsfbox{SIMULS/aNmin51-.ps} \nonumber
%\end{eqnarray}
\caption[]{
We now take into account the Higgs longitudinal perturbations, and show
the evolution of the metric perturbations 
($|\Phi_1|$, $|\Phi_2|$, $|\Phi_3|$),
for (top) a long wavelength mode, crossing the Hubble radius
at $N=-53$, and (bottom) a small wavelength mode, crossing at $N=-43$.
}
\label{fig+}
\end{figure}

\begin{figure}[p]
$$
\epsfxsize=14cm \epsfbox{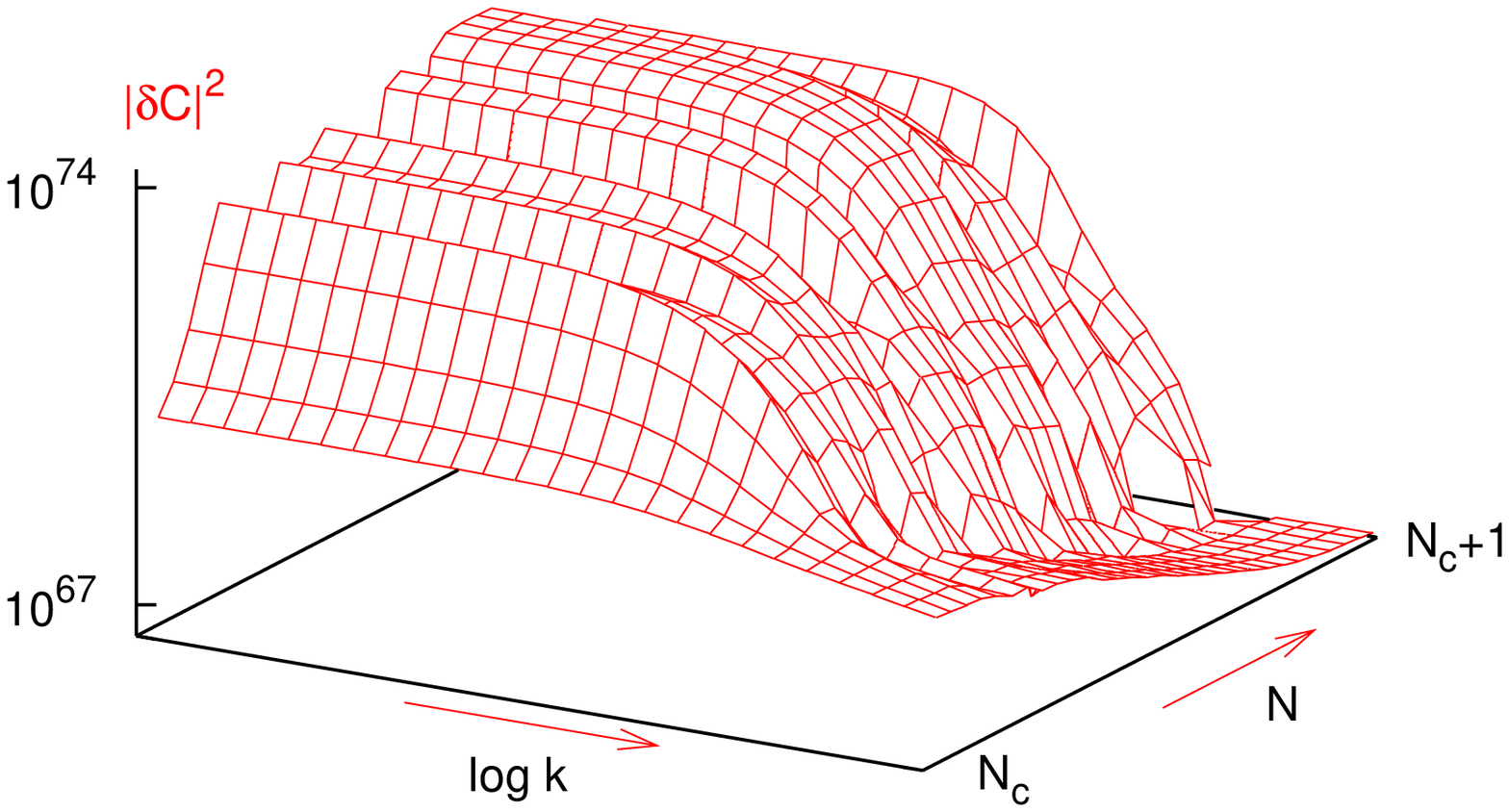}
$$
\vspace{-2cm}
$$
\epsfxsize=14cm \epsfbox{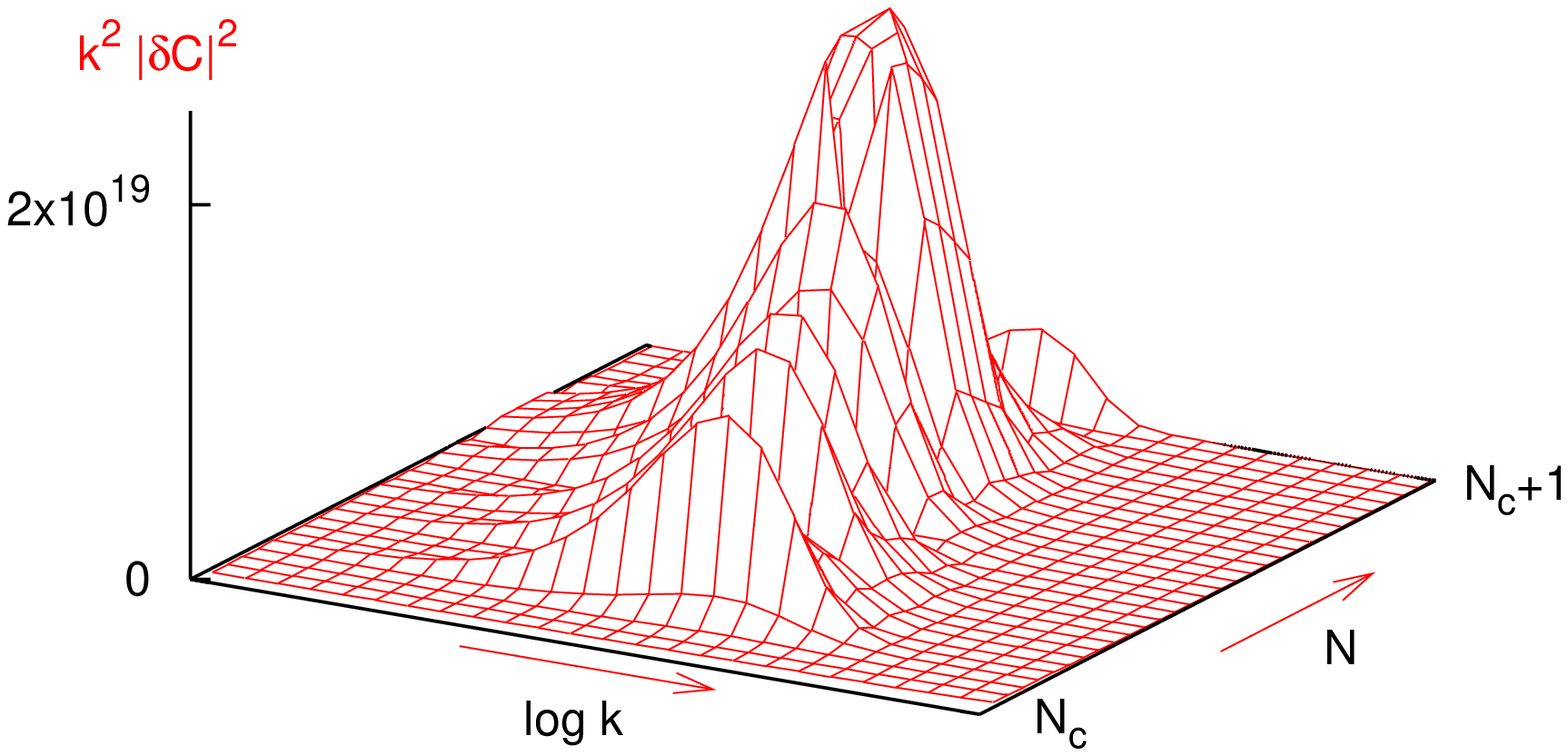}
$$
\caption[]{Before calculating the expectation value of quantum
perturbations, and compare it with background quantities, it is useful
to follow the various power spectra, plotted here for $\delta C$. The
upper plot, $|\delta C|^2$ (logarithmic scale) versus $k$ and $N$,
shows that spinodal modes are exponentially amplified, and then
oscillate coherently. The second plot, $k^2 |\delta C|^2$ (linear
scale) versus $k$ and $N$, shows that the main contribution to the
averaging integral arises always from the same modes, with $k \sim
k_{max}$; this plot shows how to define the limits of integration in
the averaging integral.}
\label{fig3D}
\end{figure}

\begin{figure}[p]
$$
\epsfxsize=12cm \epsfbox{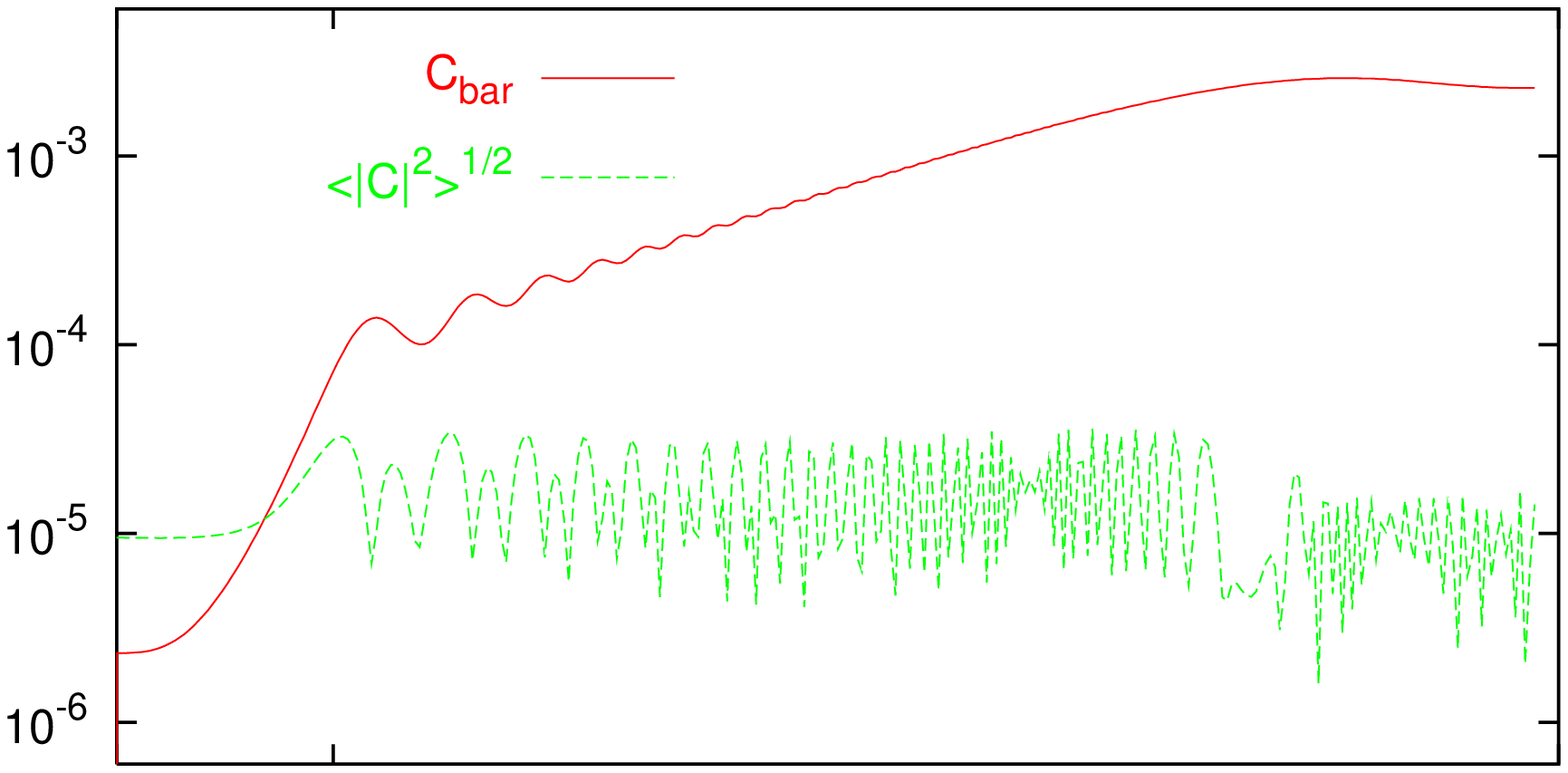}
$$
\vspace{-1.5cm}
$$
\epsfxsize=12cm \epsfbox{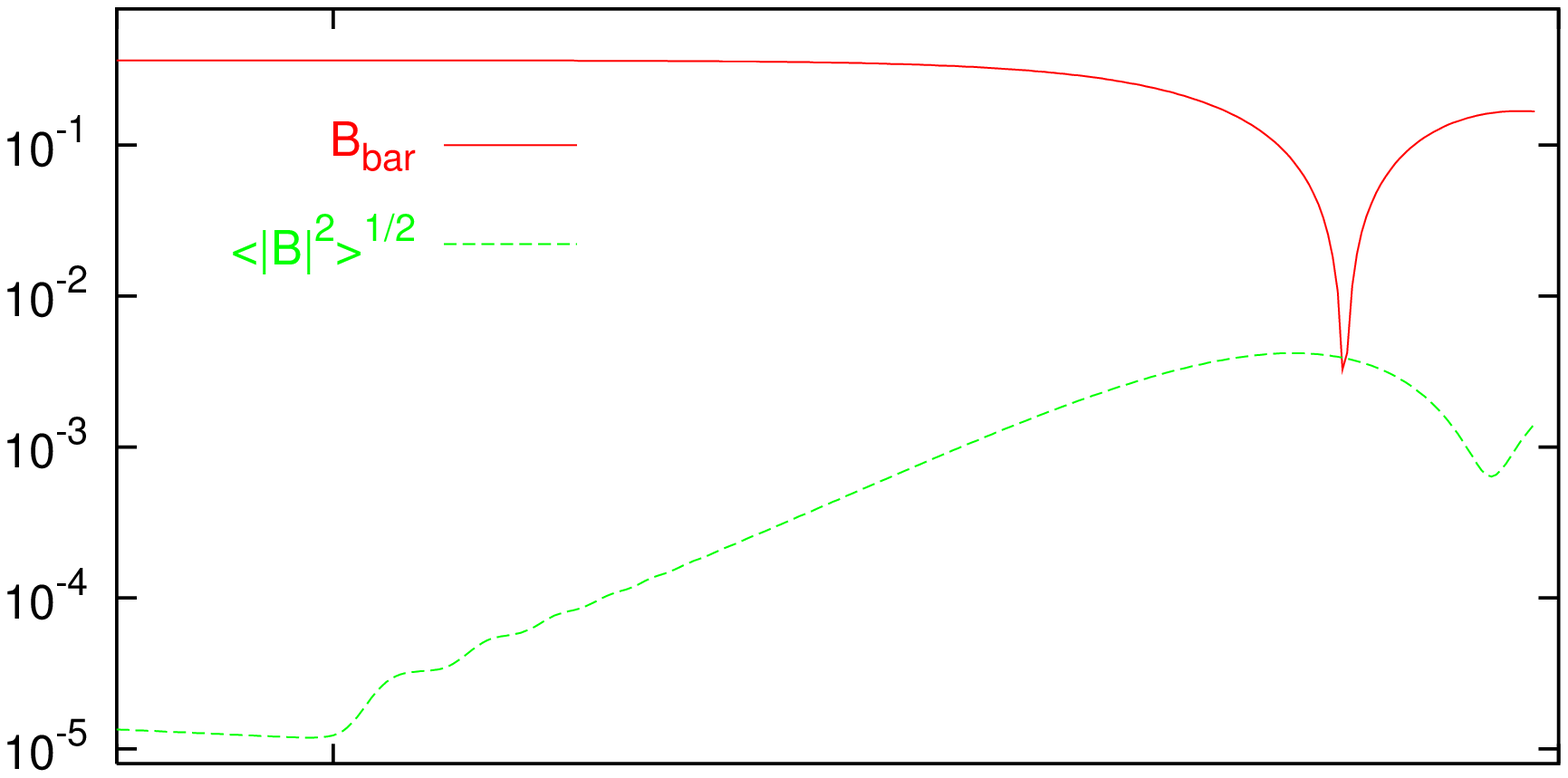} 
$$
\vspace{-1.5cm}
$$
\epsfxsize=12cm \epsfbox{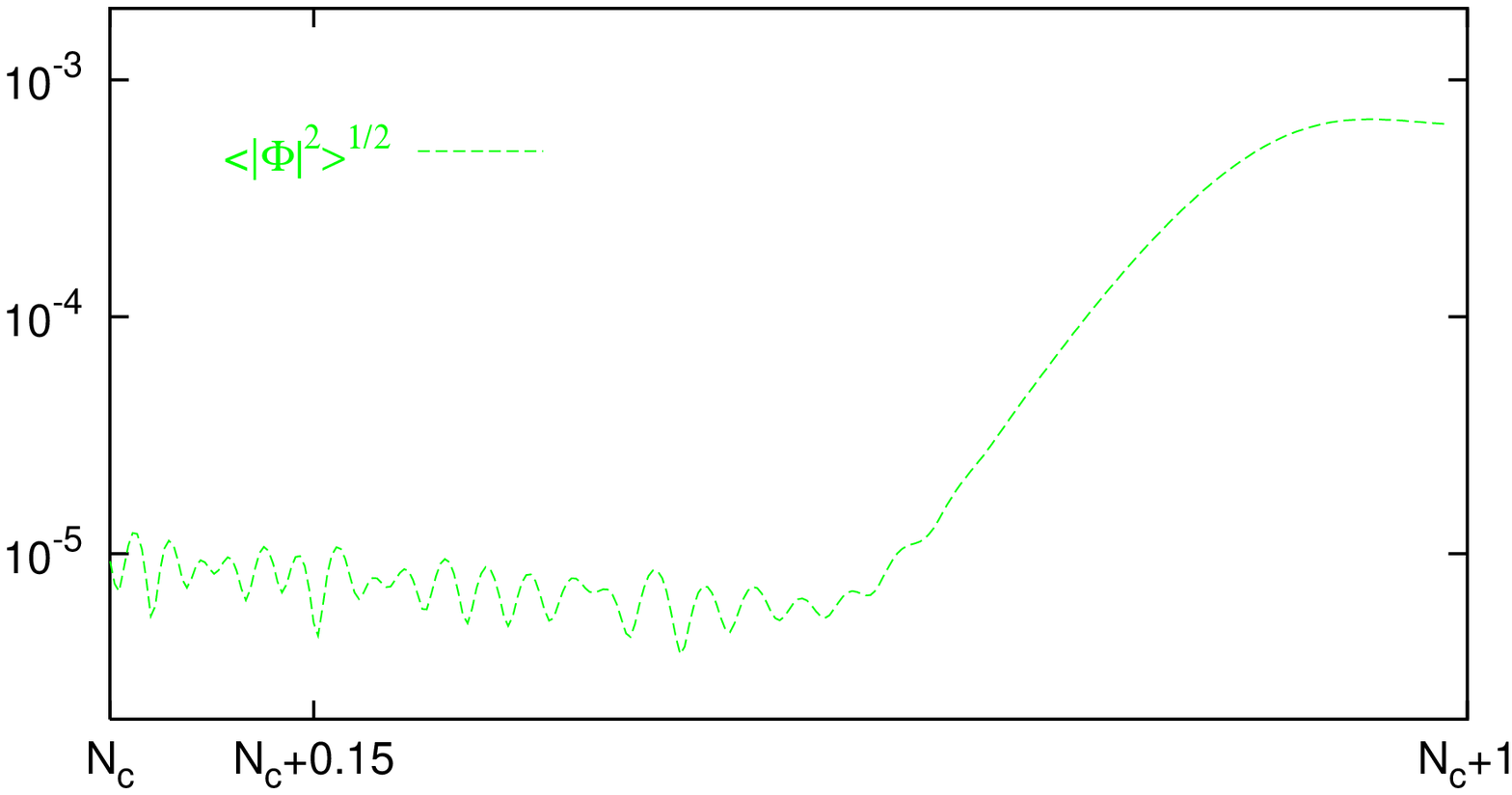} 
$$

\caption[]{Consistency check of the semi-classical approximation. The
expectation value of the quantum perturbations
$<\!|\varphi|^2\!>^{1/2}$ is generally much smaller than the
background fields $\bar{\varphi}$ (than one in the case of metric
perturbations). When it is not the case, both quantities are very
small and the linear approximation is still valid. The three plots
correspond to the case of $C$, $B$, and $\Phi$. The case of $A$ is not
shown because we know from the beginning that this field remains
extremely homogeneous. The evolution is shown only until $N=N_c+1$,
which is the time of maximal inhomogeneity, as can be seen on
fig.\ref{fig+}.}
\label{figCCH}
\end{figure}

\begin{figure}[p]
$$
\epsfxsize=10cm \epsfbox{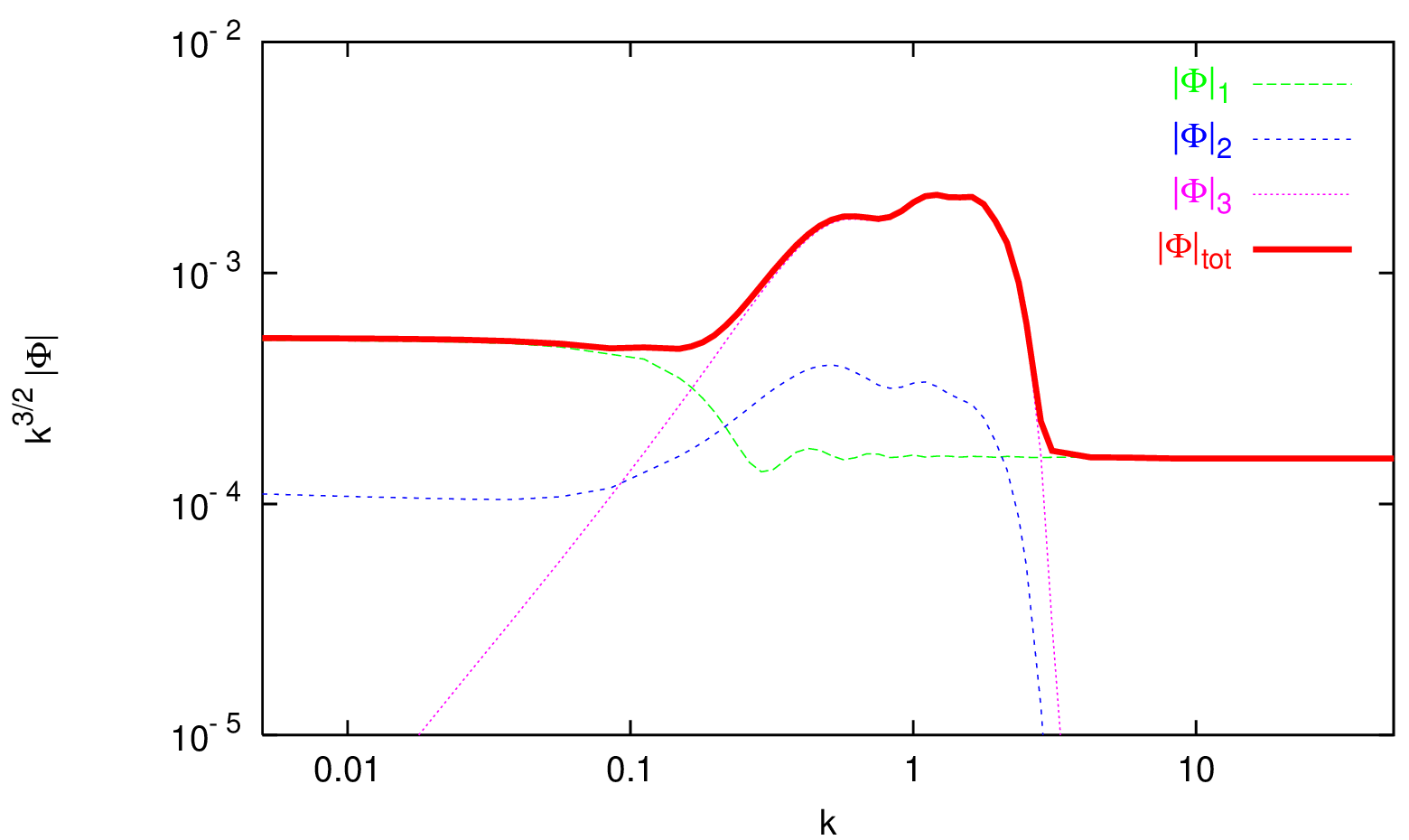}
$$
$$
\epsfxsize=10cm \epsfbox{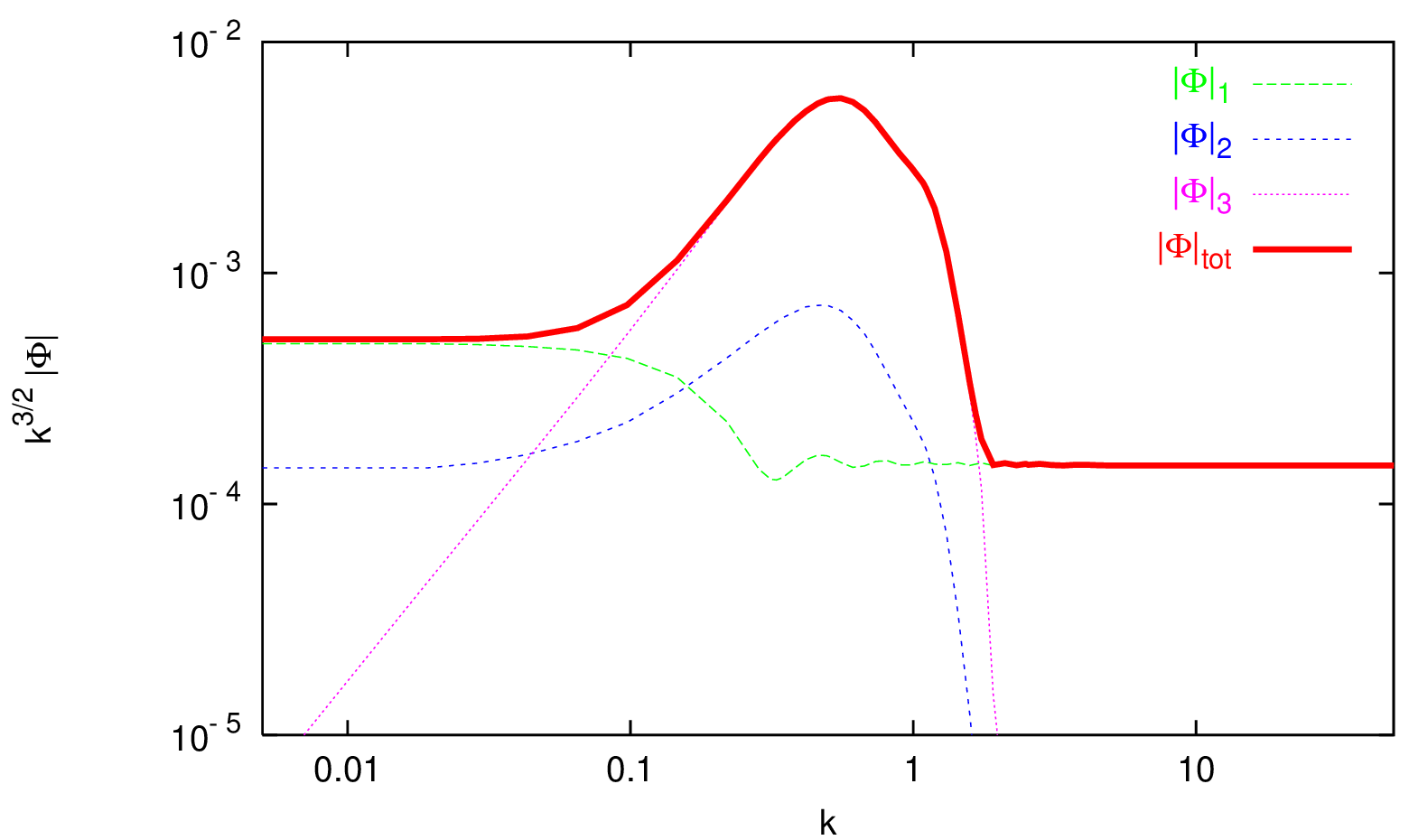}
$$
$$
\epsfxsize=10cm \epsfbox{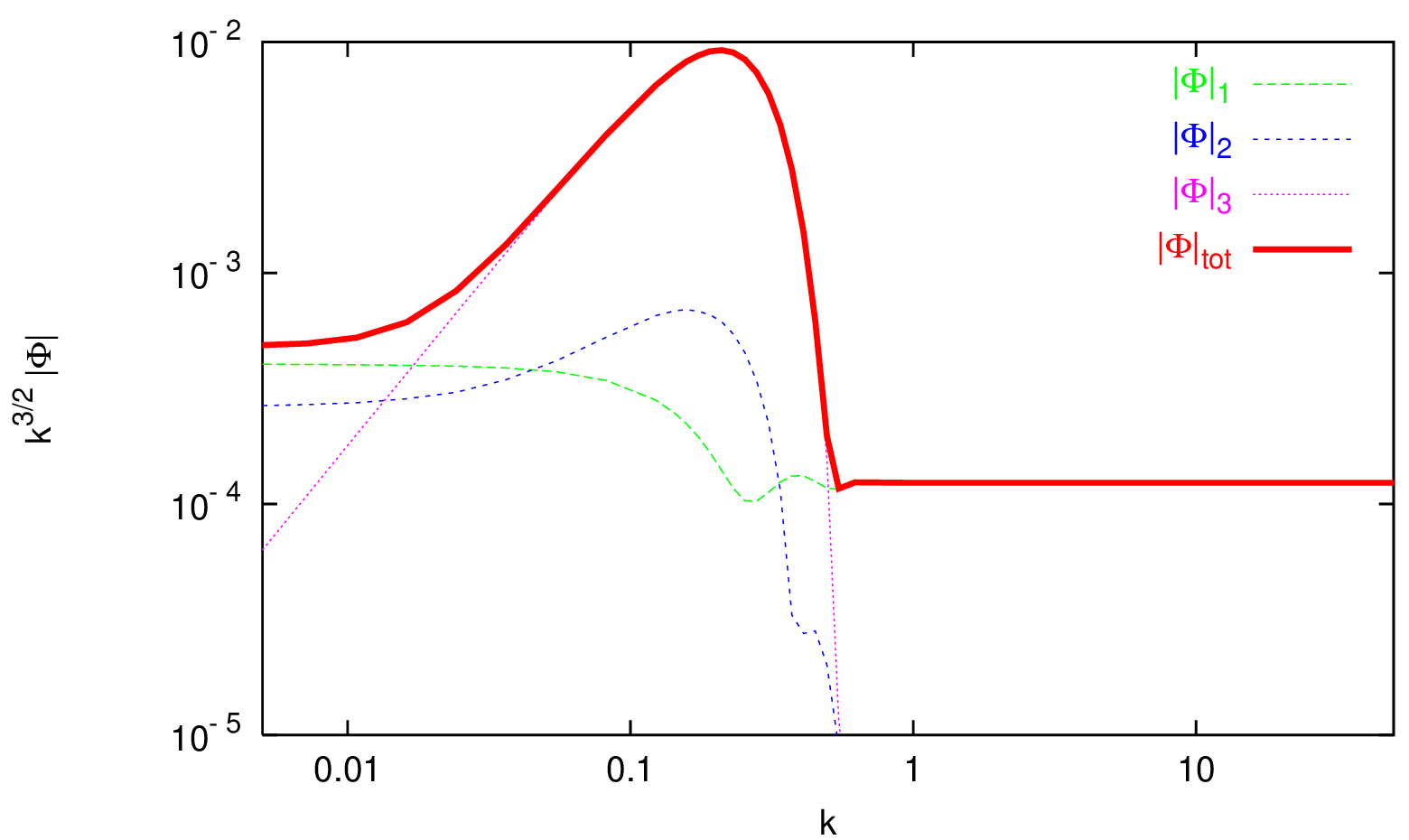}
$$
\caption[]{The primordial power spectrum of adiabatic fluctuations,
$k^{3/2} {\cal C}(k)$, or equivalently, $k^{3/2} \Phi(k)$, when the
Higgs longitudinal perturbations are taken into account (in
logarithmic scale and arbitrary units). Together with the total power
spectrum, we plot the three contributions from $\Phi_{j=1,2,3}$.  For
scales crossing the Hubble length during the phase transition,
$\Phi_3$ produces a spike (possibly non-Gaussian).  On the
first three plots, the parameters $(\xi_A, \xi_B, g_A, g_B)$ are fixed
to the values of the appendix, and $\beta = 2 \times 10^{-3}$ (top), $10^{-3}$ 
(middle), $0.5 \times 10^{-3}$ (bottom).}
\label{figSP+}
\end{figure}

\begin{figure}[p]
$$
\epsfxsize=10cm \epsfbox{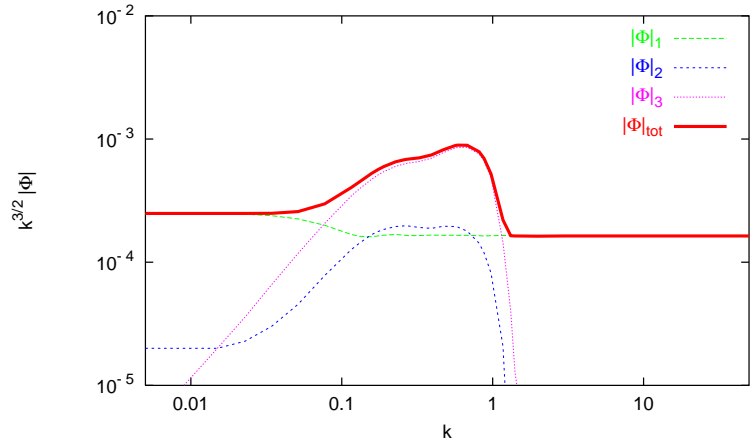}
$$
\caption[]{The primordial power spectrum of adiabatic fluctuations,
for the same parameters as on fig.\ref{figSP+} (upper plot), excepted
lower values of $\xi_A$, $\xi_B$. On the right, the small scale plateau 
is unchanged, because we kept $\xi_A-\xi_B$ fixed. On the left,
we get less power on large and intermediate scales.}
\label{figSP++}
\end{figure}

\end{document}